\documentclass[12pt]{article}
\usepackage{amssymb}
\usepackage{a4}
\def\b#1{{\mathbb #1}}
\def\c#1{{\cal #1}}

\def\nn{\nonumber \\}

\def\1{{\bf 1}}
\def\0{{\bf 0}}

\def\RH{\mbox{$\hat {\sf R}$\,}}
\def\PH{\mbox{$\sf P\!\!$}}
\def\P{\mbox{$\cal P\!\!$}}

\def\x{\mbox{x}}

\newcommand{\tl}{\,\triangleleft}

\def\cocross{{>\!\!\!\triangleleft\,}}

\def\A{\mbox{$\cal A$}}
\def\id{\mbox{id\,}}

\newcommand{\uot}{\underline{\otimes}}
\newcommand{\und}{\underline}

\newcommand{\sect}[1]{\setcounter{equation}{0}\section{#1}}

\newcommand{\app}[1]{\setcounter{section}{0}
\setcounter{equation}{0} \renewcommand{\thesection}
{\Alph{section}}\section{#1}}

\newcommand{\be}{\begin{equation}}
\newcommand{\ee}{\end{equation}}
\newcommand{\bea}{\begin{eqnarray}}
\newcommand{\eea}{\end{eqnarray}}
\newcommand{\ba}{\begin{array}}
\newcommand{\ea}{\end{array}}
\newtheorem{prop}{Proposition}
\newtheorem{lemma}{Lemma}
\newtheorem{theorem}{Theorem}

\def\sq{\mbox{\rlap{$\sqcap$}$\sqcup$}}
\newenvironment{proof}[1]{\vspace{5pt}\noindent{\bf Proof #1}\hspace{6pt}}%
{\hfill\sq}
\newcommand{\bp}{\begin{proof}}
\newcommand{\ep}{\end{proof}\par\vspace{10pt}\noindent}
\begin{document}

\title{$q$-Quaternions and $q$-deformed  $su(2)$ instantons}

\author{        Gaetano Fiore, \\\\
         \and
        Dip. di Matematica e Applicazioni, Universit\`a ``Federico II''\\
        V. Claudio 21, 80125 Napoli, Italy\\
        \and
        \and
        I.N.F.N., Sezione di Napoli,\\
        Complesso MSA, V. Cintia, 80126 Napoli, Italy
        }
\date{}

\maketitle \abstract{We construct (anti)instanton solutions of a
would-be $q$-deformed $su(2)$ Yang-Mills theory on the quantum Euclidean
space $\b{R}_q^4$ [the $SO_q(4)$-covariant noncommutative space]
by reinterpreting the function algebra on the latter as a $q$-quaternion
bialgebra. Since the (anti)selfduality equations are covariant under the
quantum group of deformed rotations, translations and
scale change, by applying the latter
we can generate new solutions from the one centered at the origin
and with unit size.  We also construct multi-instanton
solutions. As they depend on noncommuting
parameters playing the roles of `sizes' and `coordinates of the centers' of the
instantons, this indicates that the moduli space of a complete theory should be
a noncommutative manifold. Similarly, gauge transformations should
be allowed to depend on additional noncommutative parameters.
}

\vfill
\newpage

\sect{Introduction}

The search for instantonic solutions  has become a
central point of investigation of Yang-Mills gauge theories on noncommutative
manifolds after the discovery \cite{NekSch98}  that deforming
$\b{R}^4$ into the Moyal-Weyl noncommutative Euclidean space
 regularizes the zero-size singularities of the
instanton moduli space (see also \cite{SeiWit99}). Other noncommutative
gemetries have been considered, mostly deformations
\cite{ConLan01,BonCicTar00,DabLanMas01,LanvSu06} of the sphere $S^4$,
because the latter, as a compactification of $\b{R}^4$, provides a better
framework to display the topological properties of the instanton bundles.
It is therefore tempting to investigate this issue also on another available
deformation of $\b{R}^4$, the
Faddeev-Reshetikhin-Takhtadjan noncommutative Euclidean space $\b{R}_q^4$
covariant under $SO_q(4)$ \cite{FadResTak89}.

At least in our opinion, there is still no
fully satisfactory formulation  of gauge field theory on quantum group
covariant noncommutative spaces (shortly: quantum spaces) like $\b{R}_q^4$
(see e.g. \cite{JurMoeSchSchWes01} for an attempt). One main reason is the
lack of a proper (i.e. cyclic) trace to define gauge-invariant observables
(action, etc). Another one is the $\star$-structure of the
differential calculus, which for real $q$ is problematic. Probably a
satisfactory formulation will be possible within a generalization of
the standard framework of noncommutative geometry \cite{Con94}
where gauge transformations, gauge potentials, and the corresponding field
strengths will depend not only on coordinate, but also on derivatives
(as suggested e.g. in \cite{DimMeyMoeWes03,AscDimMeySchWes06}) and/or
possibly on additional noncommuting parameters (see section \ref{shift} below).
Here we leave these issues aside and just ask for nontrivial differential
2-forms solutions of the deformed (anti)selfduality equations:
results in this direction might contribute to suggest more general
formulations of gauge theories on noncommutative manifolds that include
quantum spaces.

As known, the search and classification \cite{AtiDriHitMan78} of
Yang-Mills instantons on $\b{R}^4$ is greatly simplified
when the latter is endowed with the structure of a
quaternion algebra $\b{H}$. Therefore,
following the undeformed case, we first (section \ref{quater})
introduce a notion of a $q$-quaternion as a $2\times 2$ matrix which can
be factorized as the product of the
defining matrix of $SU_q(2)$ by an element of a semigroup isomorphic to
the semigroup $\b{R}^{\ge}$ of nonnegative
real numbers,
and reformulate
the algebra $\A$ of functions on $\b{R}_q^4$ as a $\star$-bialgebra
$C\big(\b{H}_q\big)$.  The bialgebra structure encodes the property that the
product of two quaternions is a quaternion and is inherited from the bialgebra of
$2\times 2$ quantum matrices \cite{Dri86,FadTak86,Wor87a,FadResTak89}
(therefore it differs from the proposal in \cite{Maj94}).
We shall give more details and further developments in Ref.  \cite{Fio05q}.
It also turns out that our $\star$-algebra $\A$ and the
$C^{\star}$-algebra of functions on the quantum 4-sphere of Ref.
\cite{DabLanMas01} are made to be isomorphic (as $\star$-algebras) if they are
slightly extended so as to contain suitable rational functions of their
respective central elements; therefore that noncommutative sphere can be
regarded as a compactification of $\A$. In
section \ref{covdiffcal} we reformulate in $q$-quaternion language  the
$SO_q(4)$-covariant differential calculus [this turns out  to coincide
with the bicovariant differential calculus on $M_q(2),GL_q(2)$
\cite{S2,SchWatZum92}, and after imposing the unit $q$-determinant condition with
the Woronowicz 4D- bicovariant differential calculus \cite{Wor89,PodWor90} on
$C\left(SU_q(2)\right)]$, the $SO_q(4)$-covariant $q$-epsilon tensor and Hodge
map \cite{Fio94,fiothesis,Maj95jmp,Fio04JPA} on $\Omega^*(\b{R}^4_q)$. In
section \ref{gaugeth} we recall some basic notions about the standard framework
\cite{Con94} for gauge theories on noncommutative spaces, pointing out where
it doesn't fit the present model, and we formulate (anti)selfduality
equations. In section \ref{instconst} we find a large family of solutions $A$
of the (anti)self-duality equations in the form of 1-form valued $2\times 2$
matrices both in the ``regular'' and in the ``singular gauge''. There is a
larger indeterminacy than in the undeformed theory because we are not yet able
to formulate and impose the correct antihermiticity condition on the gauge
potential. Among the solutions there are some distinguished choices that
 closely resemble (in $q$-quaternion language) their undeformed counterparts (instantons and anti-instantons) in
$su(2)$ Yang-Mills theory on $\b{R}^4$. The (still missing)
complete gauge theory might however be  a deformed $u(2)$ rather
than $su(2)$ Yang-Mills theory.
We also make contact with the today standard formulation \cite{Con94} of gauge
theory on noncommutative spaces based on the identification
of vector bundles on the latter with projective modules over $\A$
by constructing in $q$-quaternion language the hermitean projector
associated to the $q$-deformed instanton projective module, and
we find that it coincides (for a specific choice of the
instanton size parameter) with the one found in Ref. \cite{DabLanMas01}.
As in the
undeformed (and in the Nekrasov-Schwarz \cite{NekSch98}) case,
applying (section \ref{shift}) the quantum group $SO_q(4)$
of $q$-deformed rotations  one obtains  gauge
equivalent solutions (by a global gauge transformation),
whereas applying that of $q$-deformed dilatations and the braided group
of $q$-deformed translations one finds
gauge inequivalent solutions. The difference is however that
a dependence on additional noncommutative parameters is introduced:
this global gauge transformation depends
the noncommuting coordinates of $SO_q(4)$, whereas the gauge inequivalent solutions
depend on the noncommuting ``coordinates of the center''
of the (anti)instanton.
Finally (section \ref{multiinst}), we find  first $n$-instantons solutions
 in the ``singular'' gauge for any integer $n$; the construction procedure
is not yet the deformed analog of the general ADHM one\cite{AtiDriHitMan78},
but rather of the procedure initiated in \cite{tHo76} and developed
in \cite{vari}, which reduces to the determination of a suitable scalar
potential, expressed in quaternion language. Then for $n=1,2$ we transform
the singular solutions into ``regular'' solutions by ``singular gauge
transformations'', as in the undeformed case (of course the $n=1$ regular
instanton solution is again the one found in section \ref{gaugeth}). The solutions
are parametrized by noncommuting parameters playing the role of ``sizes'' and
``coordinates of the centers'' of the (anti)instantons. This indicates that
the moduli space of a complete theory will be a noncommutative manifold.

\sect{Promoting $C\left(\b{R}^4_q\right)$ to the $q$-quaternion
bi- (or Hopf) algebra $C\left(\b{H}_q\right)$}
\label{quater}

We start by recalling how the (undeformed) quaternion
$\star$-algebra
$\b{H}$
can be formulated in terms of $2\times 2$ matrices:
any $X\in \b{H}$ is given by
\[
X=\x_1+\x_2i+\x_3j+\x_4k,
\]
with $\x\in \b{R}^4$ and imaginary $i,j,k$ fulfilling
\[
i^2=j^2=k^2=-1,\qquad\qquad  ijk=-1.
\]
One refers to $\x_1$ and to the following three terms as to the `real'
and `imaginary' part of $X$ respectively.
Replacing $i,j,k$ by Pauli matrices times the imaginary unit ${\rm
i}$ we can associate to $X$ a matrix
\[ X\leftrightarrow x\equiv\left(\ba{cc}
\x_1+\x_4{\rm i} \:& \x_3+\x_2{\rm i}\\
-\x_3+\x_2{\rm i} \:& \x_1-\x_4{\rm i} \ea\right)=: \left(\ba{cc}
\alpha\:& -\gamma^{\star}\\
\gamma\:& \alpha^{\star} \ea\right) \] (where $\alpha,
\gamma\in\b{C}$).
The quaternionic product becomes represented
by matrix multiplication, and the  quaternionic
conjugation becomes represented
by hermitean conjugation of the matrix $x$.
  Therefore  $\b{H}$ can be seen also
as the subalgebra of $M_2(\b{C})$ consisting
of all complex $2\times 2$ matrices of this
form. Since the determinant of any $x$ is nonnegative,
$$
|x|^2\equiv\det(x)=|a|^2+|\gamma|^2 \ge 0,
$$
any $x$ can be factorized in the form
$$
x=T|x|,
$$
where $T\in SU(2)$ and $|x|$ belongs
to the semigroup $\b{R}^{\ge}$ of nonnegative real numbers.
Hence any $x$ belongs also to the
semigroup  $SU(2)\times \b{R}^{\ge}$.

\medskip
We $q$-deform this just replacing $SU(2)$
by $SU_q(2)$ in the dual picture of the algebra
of functions of the matrix elements of $x$. In other words, we define a
$q$-quaternion just as one introduces the defining matrix of
$SU_q(2)$ \cite{Wor87,Wor87a}, but without imposing the
unit $q$-determinant condition.
For %
$q\in\b{R}\setminus \{0\}$
consider the unital associative
$\star$-algebra $\A\equiv C(\b{H}_q)$ generated by elements $\alpha,
\gamma^{\star},\alpha^{\star},\gamma$ fulfilling the commutation
relations
\be
\ba{l}
\alpha\gamma=q\gamma\alpha,\qquad\alpha\gamma^{\star}=q\gamma^{\star}\alpha,
\qquad\gamma\alpha^{\star}=q\alpha^{\star}\gamma,\\[8pt]
\gamma^{\star}\alpha^{\star}=q\alpha^{\star}\gamma^{\star},\qquad
[\alpha,\alpha^{\star}]=(1\!-\!q^2)\gamma\gamma^{\star} \qquad
[\gamma^{\star},\gamma]=0.\ea \label{explqquatcomrel}
\ee
Introducing the matrix
\be
x\equiv \left(\ba{ll} x^{11}\: & x^{12}\\
x^{21}\:& x^{22}\ea\right):=\left(\ba{cc} \alpha\:& -q\gamma^{\star}\\
\gamma\: & \alpha^{\star} \ea\right)
\ee
we can rewrite these commutation relations as
\be
\hat R x_1x_2=x_1x_2\hat R
      \label{qquatcomrel}
\ee and the conjugation relations as
$x^{\alpha\beta}{}^{\star}=\epsilon^{\beta\gamma}x^{\delta\gamma}
\epsilon_{\delta\alpha}$, i.e.
\be
x^{\dagger}=\bar x\qquad\qquad
\mbox{where } \bar a:=\epsilon a^T\epsilon^{-1} \quad\forall a\in
M_2. \label{qquatstarrel}
\ee
Here as usual $x_1\equiv
x\otimes_{\b{C}} I_2$, $x_2\equiv I_2\otimes_{\b{C}} x$ ($I_2$ is
the $2\times 2$ unit matrix), $\hat R$ is the braid matrix of
$M_q(2)$, $GL_q(2)$ and $SU_q(2)$ \be \hat
R^{\alpha\beta}_{\gamma\delta}=q\delta^{\alpha}_{\gamma}
\delta^{\beta}_{\delta}+\epsilon^{\alpha\beta}\epsilon_{\gamma\delta},
                                               \label{Rexpl}
\ee
and $\epsilon$ is the corresponding completely
$q$-antisymmetric
tensor
\be
\epsilon\equiv (\epsilon_{\alpha\beta})
:=\left(\ba{cc} 0\: & 1\\ -q \: & 0
\ea\right),
\qquad\qquad
\epsilon^{-1}\equiv (\epsilon^{\alpha\beta})
=-q^{-1}(\epsilon_{\alpha\beta}).      \label{qepsilon}
\ee
So $\A:=C(\b{H}_q)$ can be naturally endowed with
a $\star$-bialgebra  structure
(we are not excluding ${\bf 0}_2$ from the spectrum of $x$),
more precisely the above real section
 of the bialgebra $C\left(M_q(2)\right)$ of
$2\times 2$ quantum matrices \cite{Dri86,FadTak86,Wor87a,FadResTak89}.
In the sequel we shall write the corresponding coproduct
\be
\Delta(x^{\alpha\gamma})=x^{\alpha\beta}\otimes x^{\beta\gamma}
                                        \label{coprod}
\ee
in the more compact matrix product form
\be
\Delta(x)=ax
\ee
where we have renamed $x\otimes \1\to a$, $\1\otimes x\to x$.
Since  the coproduct
is a $\star$-algebra map, $\Delta(x)$, or equivalently the matrix product $ax$ of any
two matrices $a,x$ with mutually commuting entries and fulfilling
(\ref{qquatcomrel}-\ref{qquatstarrel}), again fulfills the latter.
Therefore we shall call any such matrix $x$ a {\it $q$-quaternion},
and $\A:=C(\b{H}_q)$ the $q$-quaternion bialgebra. Note that, according
to this definition, the unit matrix is a $q$-quaternion.
Note that $I_2$ is a $q$-quaternion, and $x$ is a $q$-quaternion
iff $-x$ is.

As well-known, the socalled
`$q$-determinant' of $x$ \bea |x|^2
&\equiv&\det{}_q(x):=x^{11}x^{22}-qx^{12}x^{21}=\alpha^{\star}\alpha
+ \gamma^{\star}\gamma\nn &=& \frac 1{1\!+\!q^2}
x^{\alpha\alpha'}x^{\beta\beta'}
\epsilon_{\alpha\beta}\epsilon_{\alpha'\beta'}, \label{qdet} \eea
is
central, manifestly nonnegative-definite and group-like. Therefore at
representation level it will have zero eigenvalue
iff $x$ has $\0_2$ eigenvalue matrix. Replacing (\ref{Rexpl}) in
(\ref{qquatcomrel}) we find that the latter is equivalent  to
\be
x\bar x=\bar xx=|x|^2I_2.                \label{blu}
\ee

If we extend $C(\b{H}_q)$ by the new (central, positive-definite and
group-like) generator $|x|^{-1}$ (this will exclude $x=\0_2$ from the
spectrum),  the matrix $x$ becomes invertible
and we obtain even a Hopf $\star$-algebra  with antipode $S$ defined by
\be
Sx=x^{-1}=\frac{\bar x}{|x|^2}, \qquad\qquad S |x|^{-1}=|x|. \label{inverse}
\ee
The matrix elements of $T:=\frac x{|x|}$
fulfill the `RTT' \cite{FadResTak89} relations (\ref{qquatcomrel}) and
\be
T^{\dagger}=T^{-1}=\overline{T}, \qquad\qquad\det{}_q(T)=\1, \label{Tinverse}
\ee
namely generate $C\left(SU_q(2)\right)$ \cite{Wor87,Wor87a} as a
quotient subalgebra. Therefore  the $x^{\alpha\alpha'}$  generate the
(Hopf) $\star$-algebra $C\left(SU_q(2)\!\times\! GL^+(1)\right)$ of functions
on the ``quantum group $SU_q(2)\times GL^+(1)$ of nonvanishing
$q$-quaternions'' [a real section of the Hopf algebra
$C\left(GL^+_q(2)\right)$], in analogy with the $q=1$ case.

In view of the construction of instanton solutions we also extend
$\A=C(\b{H}_q)$ by adding as generators
$$
\frac 1{1+\frac{|x|^2}{\rho^2}}, \qquad\qquad \quad\rho\in\b{R}^+.
$$

\subsubsection*{Identifying $\b{H}_q$ as $\b{R}^4_q$}

One can easily verify that as a $\star$-algebra $\A:=C(\b{H}_q)$ coincides
with the algebra of functions on the $SO_q(4)$-covariant quantum Euclidean
Space $\b{R}_q^4$ of \cite{FadResTak89}. We identify the present
$qx^{11},x^{12},-qx^{21},x^{22}$ with the
generators $x^1,x^2$, $x^3,x^4$ of \cite{FadResTak89} (in their
original indices convention)
or with  the generators $x^{-2},x^{-1}$, $x^1,x^2$ in
 the convention of Ref.  \cite{Ogi92} (which has been
heavily used by the author of the present work). We shall denote by
$B\equiv(B^a_{\alpha\alpha'})$  this (diagonal and invertible) matrix entering
the linear transformation $x^a=B^a_{\alpha\alpha'}x^{\alpha\alpha'}$.
The braid matrix of $SO_q(4)$ is obtained as
\be
\RH\equiv\big(\RH^{ab}_{cd}\big)=q^{-1}\c{B}\big(\hat R\otimes_{\b{C}}
\hat R\big)\c{B}^{-1}                                \label{RH}
\ee
(recall that the tensor product of two braid matrices is again a braid matrix),
where $\c{B}^{ab}_{\alpha\beta\alpha'\beta'}:=B^a_{\alpha\alpha'}
B^b_{\beta\beta'}$. Its decomposition
\be
\RH = q\PH_s - q^{-1}\PH_A + q^{-3}\PH_t       \label{projectorR}
\ee
in orthogonal projectors follows from that of the braid matrix of
$M_q(2),GL_q(2),SU_q(2)$,
\be
\hat R=q\P_s-q^{-1}\P_a,
\ee
since $\PH:=\c{B}(\P\otimes_{\b{C}}\c{P}')\c{B}^{-1}$
is a projector whenever $\P,\c{P}'$ are\footnote{The orthonormality relations for
the $\c{P}_{\mu}$, with $\mu=s,a$,
\be
\c{P}_{\mu}\c{P}_{\nu} = \c{P}_\mu \delta_{\mu\nu}, \qquad
\sum_{\mu}\c{P}_{\mu}= I,
\ee
trivially imply the orthogonality relations for the
$\PH_{\mu}$, with $\mu=s,a,a',t$.}. In
fact,  \be
\ba{ll}
\PH_s=\c{B}(\P_s\!\otimes_{\b{C}}\!\P_s) \c{B}^{-1},
\qquad\qquad &\PH_t=\c{B}(\P_a\!\otimes_{\b{C}}\! \P_a)\c{B}^{-1},\\[8pt]
\PH_a=\c{B}(\P_s\!\otimes_{\b{C}}\! \P_a)\c{B}^{-1},
\qquad\qquad &\PH_{a'}=\c{B}(\P_a\!\otimes_{\b{C}}\! \P_s)\c{B}^{-1},\\[8pt]
\PH_A=\PH_a+\PH_{a'}.
\ea                                \label{defproj}
\ee
$\c{P}_s$, $\c{P}_a$, are respectively $GL_q(2)$-covariant
deformations of the symmetric and
antisymmetric projectors, and have dimension 3,1.
They can be expressed in terms
of the $q$-deformed $\epsilon$-tensor by
\be
\c{P}_a{}^{\alpha\beta}_{\gamma\delta}=-
\frac{\epsilon^{\alpha\beta}\epsilon_{\gamma\delta}}{q+q^{-1}},
\qquad\qquad
\c{P}_s{}^{\alpha\beta}_{\gamma\delta}=\delta^{\alpha}_{\gamma}
\delta_{\delta}^{\beta}+
\frac{\epsilon^{\alpha\beta}\epsilon_{\gamma\delta}}{q+q^{-1}}.
\label{exproj}
\ee
$\PH_s$, $\PH_A$, $\PH_t$
are $SO_q(4)$-covariant deformations of the symmetric
trace-free, antisymmetric and trace projectors respectively;
as we shall see $\PH_a,\PH_{a'}$ are projectors respectively on the
selfdual and antiselfdual 2-forms subspaces. By (\ref{defproj})
$\PH_s,\PH_a,\PH_{a'},\PH_A,\PH_t$ respectively have dimensions
9,3,3,6,1, and
\be
\PH_t{}_{kl}^{ij} = (g^{sm}g_{sm})^{-1} g^{ij}g_{kl}
= \frac1{(q+q^{-1})^2} g^{ij}g_{kl}
\label{Pt}
\ee
where the $4 \!\times\! 4$ matrix $g_{ab}$ (denoted as $C_{ab}$ in
\cite{FadResTak89}) is given by
\be
g_{ab}=B^{-1}{}^{\alpha\alpha'}_aB^{-1}{}^{\beta\beta'}_b
\epsilon_{\alpha\beta}\epsilon_{\alpha'\beta'};     \label{defg}
\ee
it is the $SO_q(4)$-isotropic 2-tensor,
deformation of the ordinary Euclidean metric, and
``Killing form'' of $U_qso(4)$.
Recalling that $\hat R^T=\hat R$
one immediately checks that the commutation relations
(\ref{qquatcomrel}) become
\be
\PH_A{}^{ij}_{kl} \, x^k x^l=0                             \label{xxrel}
\ee
as in the definition  \cite{FadResTak89} of the quantum Eulidean space.

\medskip
The commutation relations and the $\star$-structure are covariant under, i.e.
preserved by, matrix multiplication
$$
x\to a\, x\,b
$$
by the defining matrices
$a,b$ of two copies
$SU_q(2)$, $SU_q(2)'$ of the special unitary quantum
group, or of two copies $\b{H}_q$, $\b{H}_q'$ of the
quaternion quantum group, respectively, whose entries commute with each other
and with the entries of $x$. In other words they are covariant
under the (mixed left-right) coactions of $SU_q(2)\otimes SU_q(2)'=Spin_q(4)$ and
of $\b{H}_q\otimes \b{H}_q'$. This follows from the fact that the twofold coproduct
$\Delta^{(2)}(x)=axb$,  \be
\Delta^{(2)}(x^{\alpha\alpha'})=a^{\alpha\beta}b^{\beta'\alpha'}\otimes
x^{\beta\beta'},\qquad \qquad \mbox{i.e.}\quad
x\stackrel{\Delta_L}{\longrightarrow} a\, x\,b,  \label{SUq2SUq2coaction}
\ee
is a  $\star$-homomorphism, or equivalently both the
the left coaction $x\to a\, x$ and the right one $x\to x\,b$ are.

Upon applying the linear transformation $\c{B}$
(\ref{SUq2SUq2coaction}) takes the form
\be
\Delta_L(x^i)={\bf T}^i_j\otimes
x^j,\qquad \qquad  {\bf T}^i_j:= B^i_{\alpha\alpha'}
a^{\alpha\beta}b^{\beta'\alpha'}B^{-1}{}^{\beta\beta'}_j.
\label{SOq4coaction}
\ee
Note that the ${\bf T}^i_j $ are invariant under the $\b{Z}_2$ action defined
by the change of signs $(a,b)\to (-a,-b)$.
Relation (\ref{SOq4coaction})$_1$
 has the same form as the left coaction of Ref. \cite{FadResTak89}
of the quantum group $SO_q(4)$ [and of its
extension $\widetilde{SO_q(4)}:=SO_q(4)\!\times\! GL^+(1)$,
the quantum group of rotations and scale transformations in 4
dimensions] on $\b{R}_q^4$. This no formal coincidence: the
${\bf T}^i_j$ fulfill the commutation and $\star$-conjugation relations
\be
\RH{\bf T}_1{\bf T}_2={\bf T}_1{\bf T}_2\RH, \qquad\qquad
{\bf T}^i_j{}^{\star}=g^{jj'}{\bf T}^{i'}_{j'}g_{i'i}  \label{defSOq4}
\ee
and in addition $g_{ii'}{\bf T}^i_j{\bf T}^{i'}_{j'}=g_{jj'}\1$ if the central
element $|a||b|$ is 1\footnote{Relation (\ref{defSOq4})$_1$ follows from
(\ref{RH}) and (\ref{qquatcomrel}) for both $a$ and $b^T$
[note that the transpose $b^T$ also
fulfills (\ref{qquatcomrel}), as $\hat R^T=\hat R $]; relation
(\ref{defSOq4})$_2$  follows from (\ref{qquatstarrel}) for both $a$ and $b$;
$g_{ii'}{\bf T}^i_j{\bf T}^{i'}_{j'}=g_{jj'}$ follows from
(\ref{qdet}), (\ref{defg}) when $|a||b|=1$.}.   These are respectively
the  defining relations
of  $\widetilde{SO_q(4)}$ and of  the compact quantum subgroup $SO_q(4)$
\cite{FadResTak89}. We have thus an explicit realization
of the equivalences
$$
SO_q(4)=SU_q(2)\!\times\! SU_q(2)'/\b{Z}_2 ,
\qquad\quad\widetilde{SO_q(4)}=\b{H}_q\!\times\!\b{H}_q'/GL(1).
$$

As we shall recall in section \ref{shift}, the commutation relations are also
invariant  under the braided group of translations \cite{Maj92,Maj95} $\b{R}_q^4$,
which is the $q$-deformed version of the group of translations
$\b{R}^4$; the
role of composition of translations is played by the socalled braided
coaddition.
They are in fact covariant under the coaction of the full
inhomogenous extension $\widetilde{ISO_q(4)}$ \cite{SchWeiWei92} of
$\widetilde{SO_q(4)}$ (or quantum Euclidean group in 4 dimensions), which
includes $q$-deformed translations together with scale changes and rotations
($\widetilde{ISO_q(4)}$ can be obtained also by ``bosonization''
of $\b{R}_q^4$ \cite{Maj92}).

The analogy with the case $q=1$ would be complete if one were able to further extend the action of $\widetilde{ISO_q(4)}$ into that of a quantum conformal group. This is
out of the scope of this work and will hopefully treated elsewhere \cite{Fio05q}.
A quantum deformation of the Universal Enveloping Algebra (U.E.A.) of the conformal group  having the U.E.A. of
the $q$-deformed Poincar\'e group \cite{OgiSchWesZum92} as a closed subalgebra was already constructed in \cite{KobUem93}.

\subsubsection*{Comparison and links with other formulations}

A matrix version of the 4-dim quantum Euclidean
space (with no interpretation in terms of $q$-deformed quaternions)
was proposed also in \cite{Maj94}. However, the
$\star$-relations and the $SO_q(4)$-coaction are different, i.e. cannot be put
both in the form (\ref{explqquatcomrel}), (\ref{SUq2SUq2coaction}), even by a
relabelling of the generators.

\medskip
As a $\star$-algebra, our $\A$ slightly differs from the one of the quantum 4-sphere
$S_q^4$ proposed in \cite{DabLanMas01} (which was introduced as a
`suspension' of  the algebra of the quantum 3-sphere $S_q^3$), in the sense
that
a slight extension $\A^{ext}$ of  $\A$ by some rational functions of
$|x|$ contains that algebra
as a $\star$-subalgebra. Define
\bea
&&\alpha'=\sqrt{2}\alpha^{\star} \frac {2}{1\!+\!2|x|^2}e^{ia},\qquad\qquad
\alpha'{}^{\star} =\sqrt{2}\alpha\frac {2}{1\!+\!2|x|^2}e^{-ia},\nn[8pt]
&&\beta'=\sqrt{2}\gamma^{\star} \frac {2}{1\!+\!2|x|^2}e^{ib},\qquad\qquad
\beta'{}^{\star} =\sqrt{2}\gamma\frac
{2}{1\!+\!2|x|^2}e^{-ib},\qquad\label{redef} \\[8pt] && z= \frac
{1\!-\!2|x|^2}{1\!+\!2|x|^2}\nonumber
\eea
where $\alpha,\gamma,\alpha^{\star},\gamma^{\star}$ fulfill (\ref{explqquatcomrel})
and $e^{ia},e^{ib}\in U(1)$ are possible
phase factors.
Then $\alpha',\beta',z$ fulfill the defining relation (1) of the
$C^{\star}$-algebra considered in Ref. \cite{DabLanMas01}
(where these elements are respectively denoted as $\alpha,\beta,z$),
in particular
\be
\alpha'\alpha'{}^{\star}+\beta'\beta'{}^{\star}+z^2=\1,
\ee
and the invertible function $z(|x|)$ spans
$[-1,1[$, i.e. all the spectrum of $z$ except the eigenvalue $z=1$,
as $|x|$ spans all its spectrum $[0,\infty[$.
Viceversa, starting from the latter and enlarging it so that
it contains the element $(1\!+\!z)/2(1\!-\!z)=:|x|^2$
then inverting the above formulae
one obtains elements $\alpha,\gamma,\alpha^{\star}\gamma^{\star}$ fulfilling
our defining  relations (\ref{explqquatcomrel}).

The redefinitions
(\ref{redef}) have exactly the form of a stereographic
projection of $\b{R}^4$ on a sphere $S^4$ of unit radius
(recall that $x\cdot x=2|x|^2$): $S^4$ is the sphere centered at the
origin and $\b{R}^4$ the subspace $z=0$ immersing both in a $\b{R}^5$ with
coordinates defined by $X\equiv(Re(\alpha'),Im(\alpha'),
Re(\beta'),Im(\beta'),z)$.
In the commutative theory adjoining the missing point $X=(0,0,0,0,1)$ of
$S^4$ amounts to adding to $\b{R}^4$ the point at infinity, i.e.
to compactifying $\b{R}^4$  to $S^4$. We can thus regard
the transition from our algebra to the one  considered in Ref.
\cite{DabLanMas01}
as a compactification of $\b{R}_q^4$ into their $S^4_q$.

\sect{The $SO_q(4)$-covariant differential calculi}
\label{covdiffcal}

The $SO_q(4)$-covariant differential calculus \cite{CarSchWat91}
$(d,\Omega^*)$
on $\b{R}_q^4\sim\b{H}_q$ is obtained imposing covariant
homogeneous bilinear commutation relations
 (\ref{xxirel}) between the $x^i$ and the differentials
$\xi^i:=dx^i$.
 Partial derivatives are introduced through the
decomposition
$d=\xi^a\partial_a=\xi^{\alpha\alpha'}\partial_{\alpha\alpha'}$
of the ($SO_q(4)$-invariant) exterior derivative. All
other commutation relations are derived by consistency
with nilpotency and the Leibniz rule. Beside
(\ref{xxrel}), or equivalently (\ref{qquatcomrel}),  we have
\bea
&& x^h\xi^i=q\RH^{hi}_{jk}\xi^jx^k\qquad\quad \Leftrightarrow
\qquad\quad x^{\alpha\alpha'}\xi^{\beta\beta'}=
\hat R^{\alpha\beta}_{\gamma\delta}
\hat R^{\alpha'\beta'}_{\gamma'\delta'}
\xi^{\gamma\gamma'}x^{\delta\delta'},\label{xxirel}\\
&& (\PH_s\!+\!\PH_t)^{ij}_{hk}\xi^h\xi^k=0\qquad \:\Leftrightarrow
\qquad \:\P_s{}^{\alpha\beta}_{\gamma\delta}
\P_s{}^{\alpha'\beta'}_{\gamma'\delta'}\xi^{\gamma\gamma'}
\!\xi^{\delta\delta'}\!=\!0\!=\!(\xi\epsilon\xi^T)^{\gamma\delta}
\epsilon_{\gamma\delta},\qquad\quad\label{xixirel}\\ &&
\PH_A{}^{ij}_{hk}\partial_j\partial_i=0\qquad\qquad \Leftrightarrow
\qquad \partial_{\alpha\alpha'}\partial_{\beta\beta'}=
\hat R^{\delta\gamma}_{\beta\alpha}
\hat R^{-1}{}^{\delta'\gamma'}_{\beta'\alpha'}
\partial_{\gamma\gamma'}\partial_{\delta\delta'}, \label{ddrel}\\
&& \partial_i x^j \!= \!\delta^j_i\!+\!q\hat {\sf R}^{jh}_{ik} x^k\partial_h
 \quad \: \Leftrightarrow \quad \: \partial_{\alpha\alpha'}
x^{\beta\beta'}\!\!=\!\delta^{\beta}_{\alpha} \delta^{\beta'}_{\alpha'}\!+\! \hat
R^{\beta\delta}_{\alpha\gamma} \hat R^{\beta'\delta'}_{\alpha'\gamma'}
x^{\gamma\gamma'} \!\partial_{\delta\delta'}, \quad\:    \label{dxrel}\\
&& \partial^h\xi^i\!=\!q^{-1}\hat {\sf R}^{hi}_{jk}\xi^j
\partial^k\qquad \Leftrightarrow \qquad \partial_{\alpha\alpha'}
\xi^{\beta\beta'}= \hat R^{-1}{}^{\beta\delta}_{\alpha\gamma}
\hat R^{-1}{}^{\beta'\delta'}_{\alpha'\gamma'}\xi^{\gamma\gamma'}
\partial_{\delta\delta'}.\label{dxirel}
\eea
[An alternative $SO_q(4)$-covariant differential calculus
$(\hat d,\hat\Omega^*)$ is obtained replacing $q,\RH$ by
$q^{-1},\RH^{-1}$ in (\ref{xxirel}-\ref{dxirel})].
The $\xi^i$ transform under $SO_q(4)$
exactly as the $x^i$, the $\partial_i$ in the contragredient
corepresentation.
We introduce the notation
\be
\partial^{\alpha\alpha'}:=\epsilon^{\alpha\beta}
\epsilon^{\alpha'\beta'}\partial_{\beta\beta'},
\qquad\qquad \partial\equiv \big(\partial^{\alpha\alpha'}\big).
\ee
The $\partial^{\alpha\alpha'}$ fulfill the same commutation relations (among
themselves) as the $x^{\alpha\alpha'}$, and transform in the same
way under the $SO_q(4)$ coaction (equivalently,
the $\partial^a:=g^{ab}\partial_b$ commute and transform as the $x^a$). As a
consequence, the Laplacian
$\Box:=g^{hk}\partial_k\partial_h=\epsilon^{\alpha\beta}
\epsilon^{\alpha'\beta'}\partial_{\beta\beta'}\partial_{\alpha\alpha'}$ is
$SO_q(4)$-invariant and commutes with the $\partial_{\alpha\alpha'}$, and
\be
\partial\bar\partial=\bar\partial\partial=
I_2|\partial|^2\equiv I_2\frac 1{1\!+\!q^2}\Box .
\ee
From (\ref{dxrel}), (\ref{dxirel}) it follows
\bea
&&|\partial|^2 x=q^{-2}\partial
+q^2 x|\partial|^2\qquad\qquad |\partial|^2 \,\xi
=q^{-2}\xi\, |\partial|^2    \qquad          \label{utile1}\\[8pt]
&&\partial|x|^2=q^{-2}x+q^2|x|^2\partial\qquad\qquad
\partial\frac 1{|x|^2}=-q^{-4}\frac{x}{|x|^4}
+q^{-2}\frac 1{|x|^2}\partial\qquad \label{utile3}\\
&& |\partial|^2\, \frac 1{|x|^2}=  \frac {q^{-4}}{|x|^2}
|\partial|^2- \frac {q^{-6}}{|x|^4} x\cdot \partial\label{utile4}
\eea
Since the rhs of the latter formula applied to $\1$ gives zero, $1/|x|^2$ is
harmonic, as in the undeformed case.
There exists a special combination $V$ of $\1,x\cdot\partial,\Box$
which is unitary and fulfills
$$
V x^i=qx^i V,\qquad\quad
V\partial^i=q^{-1}\partial^iV, \qquad\quad V
\xi^i=\xi^iV.
$$
We add as new generator its ''inverse square root'', a unitary
element $\lambda$ such that $\lambda^2V=V \lambda^2=\1$ and
\be
\lambda x^i=q^{-1}x^i\lambda,\qquad\quad
\lambda\partial^i=q\partial^i\lambda, \qquad\quad \lambda
\xi^i=\xi^i\lambda.                        \label{Lambdaprop}
\ee

We introduce the following unital associative algebras:
\begin{itemize}

\item We shall denote by  $\bigwedge^*$  (exterior algebra, or
algebra of exterior forms)
the $\natural$-graded algebra generated
by the $\xi^i$, where the grading $\natural$ is the degree in $\xi^i$; any
component $\bigwedge^p$ having $\natural =p$ carries a
corepresentation of $SO_q(4)$ and has the same dimension
$4\choose{p}$ as in the $q=1$ case. In particular, up to a factor there
exists a unique  4-form which we shall denote as $d^4x$.
$\bigwedge^p$ is irreducible if $p\neq 2$, and, as we shall see, splits into
a selfdual and an antiselfdual part if $p=2$, exactly as in the $q=1$ case.

\item We shall denote by  ${\cal DC}^*$
(``differential calculus algebra'') the $\natural$-graded algebra generated
by $x^i,\xi^i,\partial_i$. Elements of ${\cal DC}^p$ are
differential-operator-valued $p$-forms.

\item We shall denote by $\Omega^*\equiv$ (algebra of differential forms)
the $\natural$-graded subalgebra generated by the
$\xi^i,x^i$. By definition $\Omega^0=\A$ itself, and both $\Omega^*$
and $\Omega^p$ are $\A$-bimodules. Also, we shall denote by $\Omega_S^*$ the
subalgebra and $C\big(SU_q(2)\big)$ -bimodule generated by
$T^{\alpha\alpha'}$,  $dT^{\alpha\alpha'}$ (this
is still 4-dim!), and by  $\tilde\Omega^*$
the extension of $\Omega^*$ with the unitary generators $\lambda^{\pm 1}$
obeying (\ref{Lambdaprop}).

\item We shall denote by ${\cal H}$ (Heisenberg algebra)
the subalgebra generated by the $x^i, \partial_i$. By definition, ${\cal
DC}^0={\cal H}$, and both ${\cal DC}^*$ and ${\cal DC}^p$ are ${\cal
H}$-bimodules.

\end{itemize}

{\bf Remark 1.}
The whole set of commutation relations (\ref{qquatcomrel},
(\ref{xxirel}-\ref{dxirel})  is  \cite{CerFioMad01}  in fact invariant under
the  replacement
$x^{\alpha\alpha'}/|x|^2q^2(1\!-\!q^2)\to\partial^{\alpha\alpha'}$ (this
is an algebra homomorphism).

As a corollary, on $\Omega^*$ one can
realize the action of the exterior derivative as the (graded) commutator
\be
d\omega_p=[-\theta,\omega_p\}:=-\theta\omega_p+(-)^p\omega_p\theta, \qquad\qquad
\omega_p\in\Omega^p                      \label{thetacommu}
\ee
with the special $SO_q(4)$-invariant 1-form \cite{ChuHoZum97,Ste96} (the `Dirac
Operator', in Connes' \cite{Con94} parlance)
\be
\theta:=(d|x|^2)|x|^{-2}\frac 1{q^2-1}=
\frac{q^{-2}}{q^2-1}\xi^{\alpha\alpha'}\frac{x^{\beta\beta'}}{|x|^2}
\epsilon_{\alpha\beta}\epsilon_{\alpha'\beta'}.          \label{deftheta}
\ee
$\theta$ is closed:
\be
d\theta=0,\qquad\theta^2=0.               \label{theta^2=0}
\ee
Applying $d$ to (\ref{blu})  we find
\be
x\bar\xi+\xi\bar x=(q^2\!-\!1)\theta|x|^2I_2,\qquad\qquad
\bar x\xi+\bar\xi x=(q^2\!-\!1)\theta|x|^2 I_2.   \label{xblu}
\ee
Relation (\ref{xxirel}) implies $|x|^2\xi^i=q^2\xi^i|x|^2$, which we generalize
as usual to
\be
|x|\xi^i=q\xi^i|x|,  \qquad\Rightarrow\qquad
|x|\,\theta=q\,\theta\,|x|.                 \label{theta|x|rel}
\ee
By a straightforward computation one also finds
\be
 dT^{\alpha\alpha'}=
q^{-1}\xi^{\alpha\alpha'}\frac 1{|x|}+(q^{-1}\!-\!1)\theta T^{\alpha\alpha'}.
\label{dT}
\ee

By (\ref{SUq2SUq2coaction}) the 1-form-valued $2\times 2$ matrices \be
(dT)\overline{T}, \qquad (d\overline{T})T \ee are manifestly
invariant under respectively the right and left coaction of the Hopf algebra
$SU_q(2)$, or equivalently the $SU_q(2)'$ and  the  $SU_q(2)$ part
of $SO_q(4)$ coaction. Setting
$Q:=-\epsilon^{-1}\epsilon^T$  one finds
$$
\mbox{tr}[Q(dT)\overline{T}]=\mbox{tr}[Q^{-1}(d \overline{T})T]
=(q\!-\!1)(q\!-\!q^{-2}) \theta;
$$
only in the $q\to 1$ limit these traces vanish. That's why for generic
$q\neq 1$ the four matrix elements of either $(dT)\overline{T}$ or
$(d\overline{T})T$ are independent, and make up alternative bases
for both $\Omega^*_S$ and $\Omega^*$.

Actually, one can check (we will give details in \cite{Fio06corfu})  that
$(d,\Omega^*)$ coincides
with the bicovariant differential calculus on $M_q(2),GL_q(2)$
\cite{S2,SchWatZum92}, and
$(d,\Omega^*_S)$ coincides
with the Woronowicz 4D-  bicovariant one \cite{Wor89,PodWor90} on $C\big(SU_q(2)\big)$.

\bigskip
One major problem in the present
$q\in\b{R}\setminus \{0\}$
case is that the calculus is not
real: there is no $\star$-structure such that
$d(f^{\star})= (df)^{\star}$, nor is there a $\star$-structure
$\star:\Omega^*\to\Omega^*$.
Formally, a $\star$-structure would map the commutation relations of
$(d,\Omega^*)$ into the ones of $(\hat d,\hat\Omega^*)$, and conversely.
At least, there is a $\star$-structure  \cite{OgiZum92}
$$
\star:{\cal DC}^*\to {\cal DC}^*
$$
having the desired commutative limit (the  $\star$-structure of the De Rham
calculus on $\b{R}^4$), but a rather nonlinear character (incidentally, the
latter  has been recently \cite{Fio04} recast in a much more suggestive form),
in other words objects of the second calculus can be realized nonlinearly in
terms of objects of the first (and conversely).

One could  introduce a simpler [$SO_q(4)$-covariant]
$\star$-structure \be \star":\tilde\Omega^*\to \tilde\Omega^*. \ee It
would be compactly summarized in the formula \be \theta^{\star"}=-
q\lambda^{-2}\theta, \label{startheta} \ee and would coincide with
the one suggested as a side-remark in formula (7.2) of
\cite{FioMad00}. But this would not be useful for the present purposes, because
its $q\to 1$ limit is not the $\star$-structure of the De Rham
calculus on $\b{R}^4$\footnote{Eq. (\ref{startheta}) is equivalent to
$(d|x|^2)^{\star"}=-q^{-1}\lambda^{-2}d|x|^2$. In the limit $q\to 1$
$\lambda\to \1$, so that $(d|x|^2)^{\star"}=-d|x|^2$, i.e.
$d|x|^2$ is {\it purely imaginary}, rather than real!
A short computation also shows that
in this limit $(\xi\bar\xi)^{\dagger"}
\propto T(\bar\xi\xi)T\in \Omega^2{}'$,
in other words $\star"$ maps selfdual into antiselfdual
2-forms (and conversely), instead of preserving
the chirality!}, unless in the commutative limit some coordinates
$x^a$ vanish (instead of becoming cartesian coordinates).

\subsection{Hodge operator and (anti)selfdual 2-forms}

The {\bf Hodge map}  is a $SO_q(4)$-covariant,
$\A$-bilinear map $*:\tilde\Omega^p\to\tilde\Omega^{4-p}$ such that
$*^2= \id$, defined by
\[
{}^*(\xi^{i_1}...\xi^{i_p})= c_p\,\xi^{i_{p+1}}...\xi^{i_4}
\varepsilon_{i_4...i_{p+1}}{}^{i_1...i_p}\lambda^{2p-4},
\]
where
$\varepsilon^{hijk}\equiv$ $q$-epsilon tensor \cite{Fio94,fiothesis,Maj95jmp,Fio04JPA}
and $c_p$ are suitable
normalization factors \cite{Fio94,fiothesis,Maj95jmp,Fio04JPA}.
Actually this extends \cite{Fio04JPA}
to a ${\cal H}$-bilinear
map $*:{\cal DC}^p\to{\cal DC}^{4-p}$ with the same features. For $p=2$
$\lambda$-powers disappear and one even gets a map
$*:\Omega^2\to\Omega^2$ defined by
\be
{}^*\xi^i\xi^j=\frac
1{[2]_q}\xi^h\xi^k \varepsilon_{kh}{}^{ij}\omega_{ji},
\label{defHodge2x4}
\ee
where $[2]_q=q\!+\!q^{-1}$. By an explicit calculation one finds that this amounts to
\be
{}^*\xi^i\xi^j=\left(\PH_a-\PH_{a'}\right)^{ij}_{hk}\xi^h\xi^k,
\ee
with $\PH_a,\PH_{a'}$ defined in (\ref{defproj}).
$\bigwedge^2$ splits into the direct sum
\be
{\bigwedge}^2=\check{\bigwedge}^2\oplus \check{\bigwedge}^{2}{}'
= \PH_a{\bigwedge}^2\,\oplus\,\PH_{a'}{\bigwedge}^2
\ee
of the eigenspaces $\check{\bigwedge}^2, \check{\bigwedge}^2{}'$ of $*$
with eigenvalues $1,-1$ (the ``subspaces of selfdual and antiselfdual
exterior forms''
respectively), which carry the (3,1)- and (1,3)-dimensional corepresentation
of $SU_q(2)\times SU_q'(2)$. By (\ref{defproj}), (\ref{xixirel}) and
(\ref{exproj}) $\check{\bigwedge}^2, \check{\bigwedge}^{2}{}'$ are
respectively spanned by \be
f^{\alpha\beta}:=\P_s{}^{\alpha\beta}_{\gamma\delta}
\epsilon_{\gamma'\delta'}\xi^{\gamma\gamma'}\xi^{\delta\delta'}=
\epsilon_{\gamma'\delta'}\xi^{\alpha\gamma'}\xi^{\beta\delta'}
=(\xi\epsilon\xi^T)^{\alpha\beta}\label{deff}
\ee
and their antiselfdual partner
$$
f'{}^{\alpha'\beta'}:=\P_s{}^{\alpha'\beta'}_{\gamma'\delta'}
\epsilon_{\gamma\delta}\xi^{\gamma\gamma'}\xi^{\delta\delta'}
=\epsilon_{\alpha\beta}\xi^{\alpha\alpha'}\xi^{\beta\beta'}
=(\xi^T\epsilon\xi)^{\alpha'\beta'}.  \quad
\eqno{(\ref{deff})'}
$$
As expected, only three out of the four matrix elements $f^{\alpha\beta}$
(resp. $f'{}^{\alpha'\beta'}$) are independent, as
 (\ref{xixirel}) implies
$\epsilon_{\alpha\beta}f^{\alpha\beta}=0=\epsilon_{\alpha'\beta'}f'{}^{\alpha'\beta'}$.
As a basis we can alternatively use also the matrix elements of
$\xi\bar\xi$ (resp. $\bar\xi\xi$), because
\be
(\xi\bar\xi)^{\alpha\beta}=f^{\alpha\gamma}\epsilon^{\gamma\beta},
\qquad \qquad   (\bar\xi\xi)^{\alpha'\beta'}=
\epsilon^{\alpha'\gamma'}f'{}^{\gamma'\beta'}.    \label{deff"}
\ee
From the decomposition $\PH_A=\PH_a+\PH_{a'}$ one easily finds
\be
\xi^{\alpha\alpha'}\xi^{\beta\beta'}=-\frac 1{q+q^{-1}}
[f^{\alpha\beta}\epsilon^{\alpha'\beta'}+\epsilon^{\alpha\beta}
f'{}^{\alpha'\beta'}]                 \label{xixif}
\ee

Using relations (\ref{xixirel}) and (\ref{Rexpl}) one easily
derives the  following relations
\bea
&&x^{\alpha\alpha'}f^{\beta\gamma}=
q(\hat R_{12}\hat R_{23})^{\alpha\beta\gamma}_{\lambda\mu\nu}
f^{\lambda\mu}x^{\nu\alpha'},                    \label{xfrel}\\
&&\partial^{\alpha\alpha'}f^{\beta\gamma}=q^{-1}(\hat R_{12}\hat
R_{23})^{\alpha\beta\gamma}_{\lambda\mu\nu}
f^{\lambda\mu}\partial^{\nu\alpha'}.                \label{dfrel}
\eea
The second is obtained from the first by applying $\Box$
and recalling (\ref{utile1}) (or, alternatively, Remark 1).
As done in (\ref{deff})$'$, in the sequel we shall usually label a formula
regarding antiselfdual 2-forms by
adding a prime  to the label of its selfdual counterpart,
and possibly omit it, whenever it can be obtained from the latter
by the obvious replacements. As another example, the
analog of (\ref{xfrel}) reads
$$
x^{\alpha\alpha'}f'{}^{\beta'\gamma'}=
q(\hat R_{12}\hat R_{23})^{\alpha'\beta'\gamma'}_{\lambda'\mu'\nu'}
f'{}^{\lambda'\mu'}x^{\alpha\nu'}.              \eqno{(\ref{xfrel})'}
$$
From the previous three formulae and (\ref{dfrel})$'$ it follows
that $\Omega^2$ (resp. ${\cal DC}^2$) splits into the direct sum of
$\A$- (resp. ${\cal H}$-) bimodules \be
\Omega^2=\check\Omega^2\oplus \check\Omega^{2}{}' \qquad\quad
\mbox{(resp. }{\cal DC}^2=\check{\cal DC}^2\oplus \check{\cal
DC}^{2}{}'\mbox{)}                             \label{split}
\ee
of the  eigenspaces of $*$ with eigenvalues $1,-1$ respectively,
whose elements we shall call as usual ``self-dual and anti-self-dual 2-forms''.

\begin{prop}
For any $\omega_2\in\check\Omega^2$,
$\omega_2'\in\check\Omega^2{}'$,  (resp. $\omega_2\in\check{\cal DC}^2$,
$\omega_2'\in\check{\cal DC}^2{}'$)
\be
\omega_2\,\omega_2'=\omega_2'\,\omega_2=0,        \label{ASortho}
\ee
\end{prop}
\bp~ Since $\check\Omega^2,\check\Omega^{2}{}'$ are $\A$-bimodules
(resp. $\check{\cal DC}^2, \check{\cal DC}^{2}{}'$ are ${\cal
H}$-bimodules) to prove (\ref{ASortho}) it is sufficient to prove
$$
f^{\alpha\beta}f'{}^{\gamma'\delta'}=0,\qquad\qquad
f'{}^{\gamma'\delta'}f^{\alpha\beta}=0.
$$
By construction the lhs's belong to the $(3,3)$-dimensional
(irreducible) corepresentation of $SU_q(2)\times SU_q'(2)$; at the
same time, being 4-forms, they must be proportional to the invariant
4-form $d^4x$, i.e. belong to the $(1,1)$-dimensional
corepresentation. Therefore they have to vanish. \ep



The 2-forms 
$(\xi\bar\xi)^{\alpha\beta}$, $(\bar\xi\xi)^{\alpha'\beta'}$
are exact.
One can find 1-form-valued matrices $a,a'$ such that
\be
d\,a=\xi\bar\xi, \qquad \qquad d\,a'{}=\bar\xi\xi. \label{da=xibarxi}
\ee
Clearly, they are defined up to $d$-exact terms. Among the simplest choices
we have
\be
\hat a:=-\xi\bar x, \qquad \qquad\hat a':=-\bar\xi x.
\ee
They have the following commutation relations with the coordinates:
$$
x^{\alpha\alpha'}(\hat a\epsilon)^{\beta\gamma}=
(q\hat R_{12}\hat R_{23}^{-1})^{\alpha\beta\gamma}_{\lambda\mu\nu}
(\hat a\epsilon)^{\lambda\mu} x^{\nu\alpha'}
$$
(and similarly for $\hat a'$).
The four matrix elements of $\hat a$ are all
independent and  make up an alternative basis for $\Omega^1$; they
belong to the  $(3,1)\oplus (1,1)$-dimensional (reducible) corepresentation of
$SU_q(2)\times SU_q'(2)$. (And similarly for $\hat a'$). These properties
remain true for any combination
\be
a_{\kappa}:=\hat a+\kappa\,\theta |x|^2 \,I_2
\ee
with complex $\kappa\neq \kappa_0:=q^2(q^2\!-\!1)/(q^2\!+\!1)$, whereas
there are only three independent
\be
a_{\kappa_0}{}^{\alpha\beta}={\cal
P}_s{}^{\alpha\lambda}_{\gamma\delta} (\xi\epsilon x^T)^{\gamma\delta}
\epsilon^{\beta\delta},
\label{aexplicit}
\ee
because $a_{\kappa_0}{}^{\alpha\beta}(\epsilon\epsilon^T)_{\beta\alpha}=0$;
the latter belong to the (3,1) irreducible corepresentation of
$SU_q(2)\times SU_q'(2)$. There is no other matrix $a$ with
the latter property.
In the $q=1$ limit (\ref{aexplicit}) becomes the familiar
\[
a_{\kappa_0}{}^{\alpha\beta}=-\Big(\xi\epsilon^{-1}x^T\Big)^{(\alpha\lambda)}
\epsilon^{\lambda\beta}=-\left\{Im(\xi\,\bar
x)\right\}^{\alpha\beta},
\]
where $(\alpha\lambda)$ denotes symmetrization w.r.t. $\alpha\lambda$ and $Im$
the imaginary part.

From (\ref{xblu}), (\ref{theta|x|rel}), (\ref{thetacommu}),
(\ref{theta^2=0}) we easily derive \be
a_{\kappa}a_{\kappa}=(1\!-\!\kappa)[\xi\bar\xi
+(1\!-\!q^2)\xi\theta\bar x]|x|^2=q^2(1\!-\!\kappa)[\xi\bar\xi
+(1\!-\!q^{-2})a_{\kappa}\theta]|x|^2 \ee An analogous statement
holds for their primed counterparts. By straightforward calculations
one also finds \be \overline{T} a_{\kappa}T=-
q^{-1}(1\!+\!\kappa)[\hat a'\!+\!\kappa'\theta |x|^2 \,I_2] =-
q^{-1}(1\!+\!\kappa)a'_{\kappa'} \qquad \ee where
$\kappa':=q^2/(1\!+\!\kappa)\!-\!1$. Looking for a $\kappa$ such
that $\kappa'=\kappa$ we find two solutions
$\kappa_{\pm}=-\!1\!\pm\!q$, which yield the simple changes \be \bar
T a_{\kappa_{\pm}}T=\mp a'_{\kappa_{\pm}} \ee under the `similarity'
transformation $T$; it is immediate to check that
$$
a_{\kappa_+}=-q(dT)\overline{T}|x|^2 ,
$$
which has a well defined  limit as $q\to 1$, whereas
in the same limit $a_{\kappa_-}$
diverges. Since $(\bar\xi\xi)^{\alpha'\beta'}\in \check\Omega^2{}'$,
which is a $\A$-bimodule, we also find \bea T\bar\xi\xi\overline{T}
&=& x\bar\xi\xi\frac{\bar x}{|x|^2} q^{-2}
\stackrel{(\ref{xblu})}{=}-\xi\bar x\xi\frac{\bar
x}{|x|^2}q^{-2} +(1\!-\!q^{-2})\theta|x|^2\xi\frac{\bar x}{|x|^2}
\nn &\stackrel{(\ref{xblu})}{=}& \xi\bar\xi \frac{x\bar
x}{|x|^2}q^{-2}+(q^{-2}\!-\!1)\xi\theta\bar x
+(q^2\!-\!1)\theta\xi\bar x\nn &=& \xi\bar\xi q^{-2}
+(q^{-2}\!-\!q^2)\xi\theta\bar x =\xi\bar\xi q^2
+(q^2\!-\!q^{-2})\hat a\theta \:\in\check\Omega^2{}'
\qquad \label{salabim}
\eea

\sect{Looking for a suitable noncommutative gauge theory framework}
\label{gaugeth}

We recall some minimal common elements in the formulations of  gauge theories
on commutative as well as noncommutative spaces \cite{Con94,Mad99}
 (see also \cite{Lan97,FigGraVar01}).
We denote by $\A$ the `$\star$-algebra of functions on the noncommutative
space' under consideration, by $(d,\Omega^*)$ a differential calculus
on $\A$, real in the sense that  $d(f^{\star})= (df)^{\star}$. In $U(n)$ gauge
theory the gauge transformations $U$ are unitary $\A$-valued
$n\times n$ unitary matrices, $U\!\in\! M_n(\A)\equiv
M_n(\b{C})\otimes_{\b{C}}\A$,
\be
U^{-1}=U^{\dagger}, \qquad\qquad U\in \c{U}_n.
\ee
Gauge potentials are antihermitean  $n\times n$
1-form-valued matrices $A\equiv ( A^{\dot\alpha}_{\dot\beta})$,
$A\in M_n(\Omega^1)\equiv M_n(\b{C})\otimes_{\b{C}}\Omega^1$.
The case $n=1$ corresponds to electromagnetism.
The covariant derivative $D:M_n(\Omega^p)\to M_n(\Omega^{p+1})$
is defined as usual by
\be
D\omega_p:=d\omega_p+[A,\omega_p\},
\ee
and is therefore hermitean, $D(f^{\dagger})= (Df)^{\dagger}$.
The associated field strength $F\in M_n(\Omega^2)$ is defined as usual by
\be
F:=dA+AA.                               \label{defFgen}
\ee
At the right-hand side the product $AA$ is
both a (row by column) matrix product and a wedge
product. It is automatically hermitean.
As in commutative geometry, it is immediate to prove that
$F$ satisfies the Bianchi identity
\be DF=0.
\ee
The Yang-Mills equation reads as usual
\be D{}^*F=0.
\ee
If the exterior derivative can be
realized as the graded commutator (\ref{thetacommu}) with a special
1-form \cite{Con94,Wor89,Mad99} $-\theta$, then introducing the
1-form-valued matrix $B:=-\theta I_n+A$ one finds that
\be
F=BB,
\qquad \qquad D=[B,\cdot~]
\ee
and Bianchi identity is now even more
trivial. In Connes' noncommutative geometry $-\theta$ is the socalled
'Dirac operator', which has to fulfill more stringent requirements
\cite{Con94}.

In commutative geometry the socalled Serre-Swan theorem \cite{Swa62,Con95}
states that vector bundles over a compact manifold coincide
with finitely generated projective modules
${\cal E}$ over $\A$. The gauge connection $A$ of a gauge group
(fiber bundle) acting on a vector bundle is expressed in terms
of the corresponding projector ${\cal P}$.
Therefore the projectors characterizing the projective modules
can be used to completely determine the connections.
In Connes' standard approach \cite{Con94} to noncommutative geometry
the finitely generated projective modules are the primary objects to
define and develop the gauge theory. The topological
properties of the connections can be classified in terms of topological
invariants (Chern numbers), and the latter can be computed
directly in terms of characters of the projectors (Chern-Connes
characters).

Because of the Bianchi identity, in a 4D Riemannian
geometry endowed with a (involutive) Hodge map $*$ the YM equation
is automatically satisfied by a solution of the (anti)self-duality
equations
\be \ba{ll}
{}^*F=F \qquad\qquad\qquad &\mbox{self-duality},\\
{}^*F=-F \qquad\qquad\qquad &\mbox{anti-self-duality}.
\ea                                  \label{ASDEq}
\ee
If $\Omega^2$ splits as in (\ref{split}) then $F$ is uniquely decomposed in
a selfdual and an antiselfdual part
\be
F=F^++F^-.                      \label{Fsplit}
\ee

Under a gauge transformation $U$
\be
A\to A^U=U^{-1}(AU +dU),  \qquad \Leftrightarrow    \qquad
     B\to B^U=U^{-1}BU                            \label{gaugetransf}
\ee
implying as usual
\be
F\to F^U=  U^{-1}F U.
\ee
The Bianchi identity, the  Yang-Mills equation,
the (anti)self-duality equations, the
splitting (\ref{Fsplit}), the flatness condition
$F=0$ are preserved by gauge transformations.
As usual, $A=U^{-1}dU$ implies $F=0$.

Up to normalization factors, the gauge invariant
action $S$ and the `Pontryagin index', or `second Chern number',
$\c{Q}$ (a topological invariant) are defined by
\bea
S &=& \mbox{Tr}(F\:{}^*\!F),                 \label{actionfun}\\
\c{Q} &=& \mbox{Tr}(FF),                \label{topinv}
\eea
where Tr
stands for  a positive-definite trace (as such, it has to fulfill
the cyclic property) combining the $n\times n$-matrix trace with the
integral over the noncommutative manifold. If integration $\int$
fulfills itself the cyclic property then this is obtained by simply
choosing $ \mbox{Tr}=\int \mbox{tr}$, where $\mbox{tr}$ stands for
the ordinary matrix trace. $S$ is automatically nonnegative.

$\c{Q}$ can be computed in terms of the second Chern-Connes character of the
projector ${\cal P}$ associated to the connection $A$ when
Connes' formulation of noncommutative geometry applies.

If, as in the case under discussion,  (\ref{ASortho}) holds,
$S,\c{Q}$ split into the sum, difference of the two nonnegative
contributions \bea S &=&
\mbox{Tr}(F^+\:{}^*\!F^+)+\mbox{Tr}(F^-\:{}^*\!F^-),
\label{actionsplit}
\\[8pt]
\c{Q}  &=& \mbox{Tr}(F^+\:{}^*\!F^+)-\mbox{Tr}(F^-\:{}^*\!F^-).
                                             \label{topinvsplit}
\eea
As in the commutative case, these relations imply
$S\ge |\c{Q}|$.

\medskip
In the present $\A\equiv C(\b{R}_q^4)=C(\b{H}_q)$ case the above scheme
is not fully applicable because of   two {\bf
main problems}:

\begin{enumerate}

\item Integration over $\b{R}_q^4$ fulfills a {\it deformed} cyclic property
\cite{Ste96}.

\item As already recalled, $d(f^{\star})\neq (df)^{\star}$ and there is no
$\star$-structure $\star:\Omega^*\to\Omega^*$, but only a
$\star$-structure  $\star:{\cal DC}^*\to {\cal DC}^*$ \cite{OgiZum92},
with a rather nonlinear character.

\end{enumerate}

A solution to both problems might be obtained
\begin{enumerate}

\item allowing for ${\cal DC}^1$-valued $A$ ($\Rightarrow$
${\cal DC}^2$-valued $F$'s), and/or

\item defining  a cyclic trace Tr by
$\mbox{Tr}(\omega_4):=\int\mbox{tr}\left(W^{(1)}\omega_4W^{(2)}\right)$,
with some suitable positive definite $W^{(1)}\otimes W^{(2)}\in
M_n({\cal H})\otimes M_n({\cal H}) $ (in Sweedler notation with
suppressed sum symbol). (A $W\in M_n({\cal H})$ is
a pseudo-differential-operator-valued $n\times n$
matrix).

\end{enumerate}

This hope is based on our results \cite{Fio04}: 1) the Hodge map
$*$ is not only $\A$-bilinear, but fully ${\cal H}$-bilinear; 2) the
$\star$-structure  $\star:{\cal DC}^*\to {\cal DC}^*$ can been
recast in a much more suggestive form involving only  a similarity
transformation with the realization as pseudodifferential operators
of the ribbon element $\tilde w$ and of the "vector field
generators" $\tilde Z^i_j$ of the central extension of $U_qso(4)$
with dilatations; 3)
 $d$ and the exterior coderivative $\delta:=-*d*$
become conjugated of each other $$ (\alpha_p, d\beta_{p\!-\!1}) =
(\delta\alpha_p, \beta_{p\!-\!1}), \qquad\qquad (
d\beta_{p\!-\!1},\alpha_p) = (\beta_{p\!-\!1},\delta\alpha_p) $$ if
one defines
$$
(\alpha_p, \beta_p)=\int_{\b{R}_q^4}
\alpha_p^{\star}\:\:\tilde w'{}^{1/2}\:{}^*\beta_p
$$
where $\tilde w'$ is the realization of $\tilde w$ as a
pseudodifferential operator.

\sect{$q$-deformed $su(2)$ instanton}
\label{instconst}

We look for  $A\in M_2(\Omega^1)$ solutions of the
(anti)self-duality equations (\ref{ASDEq}) virtually yielding a finite action
functional (\ref{actionfun}). Among them we expect
deformations of the (multi)instanton solutions of $su(2)$ Yang-Mills theory on
the ``commutative'' $\b{R}^4$. We first recall the instanton solution
of Belavin {\it et al.}
\cite{BelPolSchTyu75}, which we write down both in t' Hooft
\cite{tHo76} and in ADHM \cite{AtiDriHitMan78}  quaternion notation:
\bea
A &=&  dx^i\, \sigma^a\,
\underbrace{\eta^a_{ij}x^j \frac 1{\rho^2+r^2/2}}_{A^a_i},\nn
&=& -Im\left\{\xi\,\frac{\bar x}{|x|^2}\right\} \frac
1{1+{\rho^2}\frac 1{|x|^2}}\nn  &=& -(dT)\overline{T} \frac
1{1+{\rho^2}\frac 1{|x|^2}} \label{Belinst}\\
 F &=& \xi\bar\xi\,\rho^2\frac
1{(|x|^2+\rho^2)^2}. \nonumber
\eea Here $r^2:=x\cdot
x=2|x|^2$, $\sigma^a$ are the Pauli
matrices, $\eta^a_{ij}$ are the so-called t' Hooft $\eta$-symbols,
$\rho$ is the size of the instanton (here centered at the origin).
The third equality is based on the identity
$$
\xi\,\frac{\bar x}{|x|^2}=(dT)\overline{T}+I_2\frac {d|x|^2}{2|x|^2}
$$
and the observation that the first and second term at the rhs
 are respectively antihermitean and hermitean, i.e. the
imaginary and the real part of the quaternion.

\medskip
In terms of the modified gauge potential  $B:=A-\theta I_2$ a
natural Ansatz for the deformed instanton solution in the `regular
gauge' is (in matrix notation)
\be
B=\xi\frac {\bar x}{|x|^2}\,l+\theta\, I_2\,m,    \label{BAnsatz}
\ee
where $l,m$ are functions of $x$ only through $|x|$. For any $f(x)$ we shall
denote $f_q(x):= f(qx)$. Using (\ref{theta|x|rel}), (\ref{theta^2=0}),
(\ref{xblu}),  (\ref{thetacommu}), (\ref{blu}) we find
\bea
F &=& B^2= \xi\frac {\bar x}{|x|^2}\,l\,\xi\frac {\bar x}{|x|^2}\,l+
\xi\frac {\bar x}{|x|^2}\,l \,\theta\,m+ \theta\,m\,\xi\frac {\bar x}{|x|^2}\,l
+\theta\,m\,\theta\,m\nn
 &=& \xi\bar x\xi\bar x\,l_q\,l\frac { q^{-2}}{|x|^4}+
\xi\bar x \theta\,l_qm\frac {q^{-2}}{|x|^2}+ \theta\xi\bar x\,m_ql\frac
1{|x|^2} +\theta^2\,m_qm\nn
&=&\xi\left[-\bar\xi x+(q^2\!-\!1)\theta|x|^2\right]\bar x\,l_q\,l\frac {
q^{-2}}{|x|^4}+ \xi[\bar\xi+ \theta\bar x]\,l_qm\frac {q^{-2}}{|x|^2}-
\xi\theta\bar x\,m_ql\frac 1{|x|^2} \nn
&=&\xi\bar\xi(m\!- \!l)\,l_q\frac {q^{-2}}{|x|^2}
+\xi\theta\bar x\left[(q^2\!-\!1)l_ql\!+\!l_qm\!- \!q^2m_q l\right] \frac
{q^{-2}}{|x|^2}. \nonumber
\eea
A sufficient condition for $F$ to be selfdual is that
the expression in the
square bracket vanishes. Setting $h:=m/l$ this amounts
to the equation $q^2h_q\!-\!h=(q^2\!-\!1)$, which is solved by
$$
m=\left[1+\bar\rho^2\frac 1{|x|^2}\right]l,
$$
where $\bar\rho^2$ is a constant, or might be a further generator
of the algebra, commuting with $\theta$.
Replacing in the expression
for $A,F$, we find a family of solutions
\be
\ba{ll}
A_l &=\xi\bar x\frac l{|x|^2}+ \theta I_2\,
\left\{1+\left[1+\bar\rho^2\frac 1{|x|^2}\right]l\right\}\\[8pt]
&=q(dT)\overline{T} l+ \theta I_2\,
\left\{1+\left[q+\bar\rho^2\frac 1{|x|^2}\right]l\right\}, \\[8pt]
F_l &= \xi\bar\xi\frac 1{|x|^2}\bar\rho^2 \frac {q^{-2}}{|x|^2}l_ql,
\ea
\label{1ista}
\ee
parametrized by the function $l$. This large (compared to the
undeformed case) freedom in the choice of the solution is due
to the fact that we have not yet imposed in $A$ the antihermiticity
condition. Actually, we don't know yet what the `right' antihermiticity
condition is: in fact, for no $l$ is $A$ antihermitean
w.r.t. the $\star$-structure \cite{OgiZum92} mentioned
in section \ref{covdiffcal}. In any case, one should check that for the
final $A$ the
resulting $F$ decreases faster than $|x|^{-2}$ at infinity, so that
the resulting action functional (\ref{actionfun}) is finite.

The second term in (\ref{1ista})$_1$
is proportional to $d |x|^2$; in the commutative limit $q=1$ it is a
connection associated to the noncompact factor $GL^+(1)$ of $\b{H}$.
In this limit the
antihermiticity condition on $A$ amounts to the vanishing
of this term and
completely determines the solution. It factors $GL^+(1)$ out
of the gauge group to leave a pure $su(2)$ gauge theory.
In the $q$-deformed case we still ignore what the `right'
$\star$- (i.e. Hermitean) structure could be, but it could well happen that
w.r.t. the latter the  second term in (\ref{1ista})$_1$ contains also a
antihermitean (i.e. imaginary) part, which would be the connection associated
to an additional $U(1)$ factor of the gauge group and which could
not be consistently disposed of. In the latter case
the associated gauge theory would necessarily be a deformed
$u(2)$ one.

For the moment we cannot solve the ambiguity, and content ourselves
with writing the solution for a couple of selected choices of $l$.
If we choose $l$ so that the second term in (\ref{1ista})$_1$ vanishes
and set $\rho^2=\bar \rho^2q^{-1}$ we obtain
\be
\ba{l}
A=-(dT)\overline{T} \frac 1{1+\rho^2\frac 1{|x|^2}}\\[8pt]
F = q^{-1}\xi\bar\xi\frac 1{q^2|x|^2+\rho^2} \rho^2
\frac 1{|x|^2+\rho^2}.
\ea
\label{lista1}
\ee
This has manifestly the desired $q\to 1$ limit (\ref{Belinst}). The second
choice,
$$
l=-\frac{1\!+\!q^2}{1\!+\!q^4}\frac
1{1\!+\!\tilde\rho^2\frac 1{|x|^2}}\qquad\qquad
\tilde\rho^2:=\frac{1\!+\!q^2}{1\!+\!q^4}\bar\rho^2,
$$
is designed in order that $A$ is proportional to the $a_{\kappa_0}$
of (\ref{aexplicit}), so that $A^{\alpha\beta}$
span the (3,1) dimensional, irreducible corepresentation of
$SU_q(2)\times SU_q'(2)$. The result is:
\be
\ba{l}
\tilde A=-\frac{1\!+\!q^2}{1\!+\!q^4}a_{\kappa_0}\frac
1{|x|^2+\tilde\rho^2}\\[8pt] \tilde F =
\frac{1\!+\!q^2}{1\!+\!q^4}\xi\bar\xi\,\frac 1{q^2|x|^2+\tilde\rho^2}
\tilde\rho^2 \frac 1{|x|^2+\tilde\rho^2}. \ea
\label{1ista2}
\ee
This also  has the desired $q\to 1$ limit (\ref{Belinst}).
If $\bar \rho^2\neq 0$, in both cases $FF$ is regular everywhere and
decreases as $1/|x|^8$ as $x\to \infty$, therefore it virtually will yield
finite action $S$ and Pontryagin index  $\c{Q}$ upon integration.

As in the undeformed case, to make the determination of
multi-instanton solutions easier it is useful to go
to the ``singular gauge''. Note that as in the $q=1$
case $T=x/|x|$ is unitary and formally not continuous at $x=0$, so it can play
the role of a `singular gauge transformation'. In fact $A$ can be
obtained through the gauge transformation $A=T(\hat
A\overline{T}+d\overline{T})$ from the ``singular'' gauge potential
\bea
\hat A &=& \overline{T}dT\frac 1{1+|x|^2\frac 1{\rho^2}} \label{hatA0}\\
&=&-(d\overline{T})T\frac 1{1+|x|^2\frac 1{\rho^2}}\nn
&\stackrel{(\ref{dT})}{=}&- \left[q^{-1}\bar\xi\frac x{|x|^2}-\frac
{q^{-3}}{q\!+\!1}
\xi^{\alpha\alpha'}\frac{x^{\beta\beta'}}{|x|^2}\epsilon_{\alpha\beta}
\epsilon_{\alpha'\beta'}\right]\frac 1{1+|x|^2\frac 1{\rho^2}}  \label{hatA}\\
\hat F &=& \overline{T}q^{-1}\xi\bar\xi\frac 1{q^2|x|^2+\rho^2} \rho^2
\frac 1{|x|^2+\rho^2} T,
\eea
which is the analog of the instanton  solution in the ``singular gauge'' found
by 't Hooft in \cite{tHo76}. By singular gauge potential it is meant that
it has a pole in $|x|=0$.
More generally, the generic solution (\ref{1ista})
can be obtained through the gauge transformation $A_l=T(\hat
A_l\overline{T}+d\overline{T})$ from
a singular solution $\hat A_l$. The latter can be obtained
also by starting from an Ansatz like
$\hat B=\bar\xi\frac { x}{|x|^2}\,\hat l+\theta\, I_2\,\hat m$,
instead of (\ref{BAnsatz}), and imposing that
the $\bar\xi\xi$ and the $\bar\xi\theta x$ term in
$\hat F=\hat B^2$ appear in a combination proportional
to (\ref{salabim}).

A straightforward computation by means of (\ref{utile3}) shows that
$\hat A$ can be expressed also in the
form
\be
\hat A= \big(\hat{\cal D}\phi\big)\phi^{-1},              \label{Dphi}
\ee
where $\hat{\cal D}$ is the first-order-differential-operator-valued
$2\times 2$ matrix obtained from the
expression in the square bracket in (\ref{hatA}) by the
replacement $x^{\alpha\alpha'}/|x|^2\to q^4\partial^{\alpha\alpha'}$,
\be
\hat{\cal D}:=q^3\bar\xi\partial-\frac {q}{q\!+\!1}d I_2, \label{Doper}
\ee
(for simplicity we are here assuming that $\rho^2$ commutes with
$\xi\partial$) and $ \phi$ is the harmonic potential
$$
\phi:=1+\rho^2\frac 1{|x|^2}, \qquad\qquad \Box\phi|=0.
$$
This is the analog of what happens in the classical case.

\bigskip

The {\bf anti-instanton solution} is obtained just by converting
unbarred into barred matrices, and conversely, as in the $q=1$ case.
For instance, from (\ref{lista1}) we obtain the anti-instanton
solution in the regular gauge \be \ba{l} A' = - (d\overline{T})T
\,\frac 1{1+{\rho^2}\frac
1{|x|^2}},\\[8pt] F' = q^{-1}\bar\xi\xi \frac 1{|x|^2+\rho^2}
\rho^2\frac 1{q^2|x|^2+\rho^2},\ea\label{qantiinst}
\ee
and for the one in the singular gauge
$\hat A'= \big(\hat{\cal D}'\phi\big)\phi^{-1}$, where \be
\hat{\cal D}':=q^3\xi\bar\partial-\frac {q}{q\!+\!1}d I_2. \label{Doper'}
\ee

\subsubsection*{Recovering the instanton projective module of
\cite{DabLanMas01}} \label{compare}

In commutative geometry
the instanton projective module ${\cal E}$ over $\A$ and the associated gauge
connection can be most easily obtained using the quaternion formalism, in the
way described e.g. in Ref. \cite{Ati79}. $\b{H}\sim\b{R}^4$
can be compactified as $P^1(\b{H})\sim S^4$. Let $(w,x)\in\b{H}^2$ be
homogenous coordinates of the latter, and choose $w=I_2$ on the chart
$\b{H}\sim\b{R}^4$. The element  $u\in\b{H}^2$ defined by
\be
u \equiv \left(\ba{c} u_1\\ u_2\ea \right)=
\left(\ba{c} I_2\\ \frac{\rho x}{|x|^2}\ea \right)
\left(1\!+\!\frac{\rho^2}{|x|^2}\right)^{-1/2}   \label{defu}
\ee
fulfills $u^{\dagger}u=I_2\1$, and the $4\times 2$ $\A$-valued  matrix
$u$ has only three independent components.
Therefore the $4\times 4$ $\A$-valued matrix
\be
{\cal P}:=uu^{\dagger}=\left(\ba{lll} I_2 & \frac{\rho \bar x}{|x|^2}\\
\frac{\rho x}{|x|^2} & \frac{\rho^2}{|x|^2}I_2 \ea \right)
\frac 1{1\!+\!\frac{\rho^2}{|x|^2}}                    \label{calP}
\ee
is a self-adjoint three-dimensional projector. It is
the projector associated in the Serre-Swan theorem
correspondence to the gauge connection (\ref{hatA0}), by the formula $\hat
A=u^{\dagger}du$.
The associated projective module ${\cal E}$ is embedded
in the free module $\A^{16}$ seen as $M_4(\A)$, and is obtained from the latter
as ${\cal E}={\cal P}M_4(\A)$.

In the present $q$-deformed setting we immediately check that
the element $u\in\b{H}_q^2$ defined by (\ref{defu}) fulfills
$u^{\dagger}u=I_2\1$ again, so that
the $4\times 2$ $\A$-valued  matrix ${\cal P}$ defined by (\ref{calP})
is hermitean and idempotent, and has only 3 independent components.
Therefore, it defines the `instanton projective module'
${\cal E}={\cal P}M_4(\A)$ also in the $q$-deformed case.
One can easily verify that ${\cal P}$ reduces to the hermitean idempotent
$e$ of \cite{DabLanMas01} if one chooses the instanton
size as $\rho=1/\sqrt{2}$ and performs the change of generators
(\ref{redef}). Therefore, interpreting the model  \cite{DabLanMas01}
as a compactification to $S_q^4$ of ours, we can use all
the results \cite{DabLanMas01} about the Chern-Connes classes of $e$.

Unfortunately in the  $q$-deformed case it is no more true that
$\hat A=u^{\dagger}du$,
essentially because the $|x|$-dependent global factor multiplying
the matrix at the rhs(\ref{calP}) does not commute with the 1-forms
of the present calculus ($|x|\xi^i=q\xi^i|x|$).

\sect{Changing the size and shifting the center of the (anti)instanton}
\label{shift}

Applying the $\widetilde{SO_q(4)}$ coaction
(\ref{SUq2SUq2coaction}) to $|x|^2,\xi\bar x,\bar\xi x$ and using
(\ref{blu}) we obtain
\bea
&&\Delta_L\left(|x|^2\right)=|x|^2|c|^2,\nn[8pt]
&&\Delta_L\left(\xi\bar x\right)=|c|^2a\, \xi\bar x \, a^{-1},
\qquad\qquad\Delta_L\left(\bar\xi x\right)=|c|^2\bar b\, \bar\xi x\,
\bar b^{-1},\nn[8pt] &&\Delta_L\left(\xi\bar\xi \right)=|c|^2a\, \xi\bar\xi\,  a^{-1},
\qquad\qquad\Delta_L\left(\bar\xi\xi \right)=|c|^2\bar b\, \bar\xi \xi\, \bar b^{-1},
\nonumber
\eea
where $|c|^2:=|a|^2|b|^2$. The result is the same also if we
consider  $|c|^2$ as an independent parameter
and choose $a,b$ with unit $q$-determinant ($|a|=|b|=\1$). If
we apply $\Delta_L$ to the instanton  gauge potentials (\ref{1ista})
we thus find
\be
\ba{l}
\Delta_L\left(A_l(\xi,x)\right)
=aA_l\big(\xi|c|,x|c|\big)a^{-1}\\[8pt]
\Delta_L\left(F_l(\xi,x)\right) = aF_l\big(\xi|c|,x|c|\big)a^{-1}.
\ea                                            \label{transfA_l}
\ee
In particular, on the gauge potential (\ref{lista1})
\be
\ba{l}
\Delta_L\left(A\right)=-a\,(dT)\overline{T}\frac
1{1+\rho'{}^2\frac 1{|x|^2}}a^{-1} \\[8pt] \Delta_L\left(F\right) =
a\,\xi\bar\xi\frac 1{q^2|x|^2+ \rho'{}^2} q^{-1}\rho'{}^2
\frac 1{|x|^2+\rho'{}^2}a^{-1} \ea \label{transfA}\ee
where we have set $\rho'{}^2:=\rho^2|c|^{-2}$.
These gauge potentials are again solutions of the self-duality equation, since
the latter is covariant under the $\widetilde{SO_q(4)}$ coaction.
The result of the $SO_q(4)$ coaction ($|a|=|b|=\1$)
can be
reabsorbed into a (global) gauge transformation (\ref{gaugetransf}), with
$U=a^{-1}$ (and similarly $U=\bar b^{-1}$  for the anti-instanton gauge
potentials), i.e. is a gauge
equivalent solution. Note that we are thus introducing
gauge transformations depending on the additional noncommuting
parameters $a,b$.  A full  $\widetilde{SO_q(4)}$ coaction ($|c|\neq 1$) instead
involves also a change of the size of the instanton, and gives an inequivalent
solution. We can thus obtain any size starting from the instanton with unit
size.

\medskip
Having built an (anti)instanton ``centered at the origin''
with arbitrary size one would like first to translate
the center
to another point $y$, then to construct
$n$-instanton solutions ``centered at points $y_{\mu}$'',
$\mu=1,2,\ldots,n$.
The appropriate framework is to replace tensor products $\otimes$
by braided tensor products $\uot$ and apply the braided
coaddition \cite{Maj95} to the covectors $x$.
This gives new (i.e. gauge inequivalent) solutions.
The braided coaddition \cite{Maj95} of the coordinates $x$ reads \be
\und{\Delta}(x)=x\uot {\bf 1}+{\bf 1}\uot x \equiv x-y, \ee where we
have renamed $x:= x\uot{\bf 1}$, $y:=-{\bf 1}\uot x$. It follows \be
\PH_A{}^{ij}_{hk}y^hy^k=0  \qquad\quad \Leftrightarrow \qquad\quad
y\bar y=\bar yy=I_2|y|^2                     \label{yyrel0} \ee Out
of the two possible braidings we choose the following one: \bea &&
y^hx^i=q\RH^{hi}_{jk}x^jy^k\qquad\quad \Leftrightarrow \qquad\quad
y^{\alpha\alpha'}x^{\beta\beta'}= \hat
R^{\alpha\beta}_{\gamma\delta} \hat
R^{\alpha'\beta'}_{\gamma'\delta'}
x^{\gamma\gamma'}y^{\delta\delta'},\nn && \partial_i y^j \!= \!q\hat
{\sf R}^{jh}_{ik} y^k\partial_h
 \qquad\quad \: \Leftrightarrow \qquad\quad \: \partial_{\alpha\alpha'}
y^{\beta\beta'}\!\!=\! \hat
R^{\beta\delta}_{\alpha\gamma} \hat R^{\beta'\delta'}_{\alpha'\gamma'}
y^{\gamma\gamma'} \!\partial_{\delta\delta'}, \qquad\:    \label{yrel}\\
&& y^h\xi^i=q\RH^{hi}_{jk}\xi^jy^k\qquad\quad \Leftrightarrow
\qquad\quad y^{\alpha\alpha'}\xi^{\beta\beta'}=
\hat R^{\alpha\beta}_{\gamma\delta}
\hat R^{\alpha'\beta'}_{\gamma'\delta'}
\xi^{\gamma\gamma'}y^{\delta\delta'},\nonumber
\eea
(the commutation relations between $y$ and $\xi$
are determined up to a `conformal factor'; we have fixed the latter
in such a way the they look exactly as the commutations
relations between $x$ and $\xi$).
As a result,
\bea
&& d\, y=y \, d, \qquad\qquad y\theta=\theta y,\\
&& (x-y)^i\xi^j=q\RH^{ij}_{hk}\xi^h(x-y)^k,\\
&& \partial_i(x-y)^j=\delta^i_j+q\RH^{jh}_{ik}(x-y)^k\partial_h,
\eea
in other words, the differential calculus is invariant under the
replacement $x\to x\!-\!y$ (i.e. under $\und{\Delta}$).
This implies that under this replacement solutions
go into solutions. Therefore the instanton solution
with ``shifted'' center $y$
will read in the regular gauge
\be
\ba{l}
A=-d\left[\frac{(x\!-\!y)}{|x\!-\!y|}\right]
\frac{(\overline{x}\!-\!\overline{y})}{|x\!-\!y|}
\frac 1{1+\rho^2\frac 1{|x-y|^2}}\\[8pt]
F = q^{-1}\xi\bar\xi\frac 1{q^2|x-y|^2+\rho^2} \rho^2
\frac 1{|x-y|^2+\rho^2}.
\ea
\label{1istsh}
\ee
and in the singular gauge
\be
\ba{l}
\hat A= \big(\hat{\cal D}\phi\big)\phi^{-1},
\qquad\qquad \phi:=1+\rho^2\frac 1{|x-y|^2},\\
\hat F =q^{-1}\frac{(\overline{x}\!-\!\overline{y})}{|x\!-\!y|}
\xi\bar\xi
\frac{(x\!-\!y)}{|x\!-\!y|}\frac 1{q^2|x-y|^2+\rho^2} \rho^2
\frac 1{|x-y|^2+\rho^2}.
\ea                                   \label{phi-Ash}
\ee
\medskip
We conclude this section by sketching how one obtains the `infinitesimal' version of
(\ref{transfA_l}), (\ref{1istsh}), i.e. transformations of the solutions under the
action of the  cross-product
$F'\cocross U_q\widetilde{so(4)}$ \cite{OgiSchWesZum92,Maj92,fiothesis,Fio95}
(i.e. the U.E.A. of the Euclidean quantum group extended
with dilatations), where $F'$ is the subalgebra of ${\cal H}$ generated
by the $\partial_{\alpha\alpha'}$.
As known, the (right) action $\tl$ of the dual Hopf algebra $H'$
of a Hopf algebra $H$ can be obtained from the (left) coaction
$\Delta_L(v)= v_{(1)}\otimes  v_{(2)}$ of the latter (in Sweedler notation)
by the rule $v\tl h'=\langle v_{(1)},h'\rangle \: v_{(2)}$
(here $\langle ~,~\rangle$ denotes the pairing between $H,H'$).
For $H'=U_qsu(2)\otimes U'_qsu(2)$ one finds in particular
\be
v^{\alpha\alpha'}\tl gg'=[\tau(g)\,v\, \tau(g')]^{\alpha\alpha'}
=\tau^{\alpha}_{\beta}(g)v^{\beta\beta'}\tau^{\beta'}_{\alpha'}(g'),
\label{fund1}
\ee
where $v\!=\!x,\partial,\xi$,
$g\!\in\! U_qsu(2)$, $g'\!\in\! U'_qsu(2)$ [$gg'=g'g$ in $U_qsu(2)\!\otimes\! U'_qsu(2)$], and $\tau$ is the
fundamental $2$-dim representation of $U_qsu(2)$\footnote{On the
FRT \cite{FadResTak89} generators
$L^{\pm}{}^{\gamma}_{\delta}$ of $U_qsu(2)$ one has
$\tau^{\alpha}_{\beta}(L^{\pm}{}^{\gamma}_{\delta})=
\hat R^{\pm 1}{}^{\gamma\alpha}_{\beta\delta}$.}.
One finds the following transformation of the instanton solution $A_l$
under $q$-rotations:
\be
A_l\tl g'=\varepsilon(g')A_l, \qquad \qquad A_l\tl g=\tau(g_{(1)})A_l\tau(S g_{(2)})
\ee
where $\varepsilon,S$ denotes the counit, antipode of $U_qsu(2),U'_qsu(2)$
and we have used Sweedler notation (with suppressed summation index)
for the coproduct $\Delta(g)=g_{(1)}\otimes g_{(2)}$.
The transformation law for the antiinstanton solution is obtained
exchanging $g$ with $g'$.

In section \ref{covdiffcal} we have introduced partial derivatives $\partial_i$
acting from the left,
as conventional. This means that the deformed Leibniz rule takes the form
$\partial_{\alpha\alpha'} (ff')=\partial_{\alpha\alpha'}(f) f'+ 
O_{\alpha\alpha'}^{\gamma\gamma'}(f) \partial_{\gamma\gamma'}(f')$, with
a suitable linear operator $O_{\alpha\alpha'}^{\gamma\gamma'}$. The generators of infinitesimal translations in the right
action $\tl$ are instead derivatives 
$\stackrel{\leftarrow}{\partial}_{\alpha\alpha'}$ acting from the right,
i.e. $ff'\!\stackrel{\leftarrow}{\partial}_{\alpha\alpha'}=f\, (f'\!\stackrel{\leftarrow}{\partial}_{\alpha\alpha'})+ 
(f\!\stackrel{\leftarrow}{\partial}_{\gamma\gamma'})\,
\tilde O_{\alpha\alpha'}^{\gamma\gamma'}(f')$. The
quickest way to determine their action on a function (or differential from) form
$\omega$ is to recall \cite{Maj93} that this is determined by the equation
$$
\omega (x\!-\!y)=\omega (x)-(\omega (x) \tl \stackrel{\leftarrow}{\partial}_{\gamma\gamma'})\,
y^{\gamma\gamma'}+ O(y^2)
$$
namely is the coefficient of the term of degree 1 in $-y^{\gamma\gamma'}$ in the expansion of
$\omega (x\!-\!y)$ in powers of $y^{\gamma\gamma'}$ (put {\it on the right} of all
$\xi,x$'s).
One thus easily finds, for instance,
\be
A^{\alpha\beta}\tl \stackrel{\leftarrow}{\partial}_{\gamma\gamma'}=
q^{-2}\!\left[A^{\alpha\beta}\frac{x^{\lambda\lambda'}}{\rho^2\!+\!|x|^2}+
\xi^{\lambda\lambda'}\frac {\delta^{\alpha\beta}}{2(\rho^2\!+\!q^2|x|^2)}\right]
\epsilon_{\lambda\gamma}\epsilon_{\lambda'\gamma'}-\xi^{\alpha\alpha'}
\frac{\epsilon_{\alpha'\gamma'}\epsilon^{\gamma\beta}}{\rho^2\!+\!q^2|x|^2}
\ee
on the instanton  gauge potentials (\ref{1ista}). The  $\stackrel{\leftarrow}{\partial}_{\alpha\alpha'}$ can
be easily realized as elements of $F'\cocross U_q\widetilde{so(4)}$,
or also of the Heisenberg algebra ${\cal H}$.

\sect{Multi-instanton solutions}
\label{multiinst}

On the basis of the latter and of the $q=1$  results \cite{tHo76,vari},
we first look for $n$-instanton solutions of the
self-duality equation in the ``singular gauge'' in the form (\ref{Dphi}).
Beside the coordinates $x^i\equiv -y_0^i$ we introduce $n$ other
coordinates $y^i_{\mu}$, $\mu=1,2,...,n$ generating as many
$\b{R}_q^4$ and braided to each other:
\be
\ba{ll}
\PH_A{}^{ij}_{hk}y_{\mu}^hy_{\mu}^k=0  \qquad\quad &\Leftrightarrow
\qquad\quad  y_{\mu}\bar y_{\mu}=\bar y_{\mu}y_{\mu}=I_2|y_{\mu}|^2\\[8pt]
y_{\nu}^hy_{\mu}^i=q\RH^{hi}_{jk}y_{\mu}^jy_{\nu}^k\qquad\quad &\Leftrightarrow
\qquad\quad y_{\nu}^{\alpha\alpha'}y_{\mu}^{\beta\beta'}=
\hat R^{\alpha\beta}_{\gamma\delta}
\hat R^{\alpha'\beta'}_{\gamma'\delta'}y_{\mu}^{\gamma\gamma'}y_{\nu}^{\delta\delta'}
\ea
\label{yyrel}
\ee
with $\mu<\nu$.
We shall call $\A_n$ the larger algebra  generated by
the $y_{\mu}^i$'s
and by  parameters $\rho_{\mu}$, $\mu=1,...,n$
fulfilling the commutation relations
\bea
&&\rho_{\nu}^2\rho_{\mu}^2=q^2\,\rho_{\mu}^2\rho_{\nu}^2,\qquad \qquad
\nu<\mu,            \label{new2}\\
&&\rho_{\nu}^2y_{\mu}^i=y_{\mu}^i\rho_{\nu}^2  \cases{q^{-2}\:\quad
\nu<\mu, \cr 1, \:\quad\nu \ge\mu .}\label{new3}
\eea
We shall also enlarge $\A_n$ to the extended Heisenberg algebra ${\cal H}_n$
and extended algebra of differential forms  $\Omega^*(\A_n)$ by adding
as generators the $\partial_i$ and the $\xi^i$ respectively, and to
the  extended differential calculus algebra ${\cal DC}(\A_n)$ by
adding as generators both the $\xi^i,\partial_i$, with cross
commutation relations \bea &&\rho_{\mu}^2\xi^i=\xi^i\rho_{\mu}^2,
\qquad\qquad\qquad\qquad\qquad
\partial_i\rho_{\mu}^2=\rho_{\mu}^2\partial_i, \label{new1}\\
&& \partial_i y_{\mu}^j \!= \!q\hat {\sf R}^{jh}_{ik} y_{\mu}^k\partial_h
 \qquad\quad \: \Leftrightarrow \qquad\quad \: \partial_{\alpha\alpha'}
y_{\mu}^{\beta\beta'}\!\!=\! \hat
R^{\beta\delta}_{\alpha\gamma} \hat R^{\beta'\delta'}_{\alpha'\gamma'}
y_{\mu}^{\gamma\gamma'} \!\partial_{\delta\delta'}, \qquad\:
\label{ymudrel}\\ && y_{\mu}^h\xi^i=q\RH^{hi}_{jk}\xi^jy_{\mu}^k\qquad\quad
\Leftrightarrow \qquad\quad y_{\mu}^{\alpha\alpha'}\xi^{\beta\beta'}=
\hat R^{\alpha\beta}_{\gamma\delta}
\hat R^{\alpha'\beta'}_{\gamma'\delta'}
\xi^{\gamma\gamma'}y_{\mu}^{\delta\delta'}, \label{ymuxirel}
\eea
Note that the first  relations, together with the decomposition
$d=\xi^i\partial_i$, imply
\be
d\,\rho_{\mu}^2=\rho_{\mu}^2 d.
\ee
Also, from these relations it is evident that
$\check\Omega^2(\A_n),\check\Omega^{2}{}'(\A_n)$ are $\A_n$-bimodules
(resp. $\check{\cal DC}^2(\A_n), \check{\cal DC}^{2}{}'(\A_n)$ are
${\cal H}_n$-bimodules).

In the sequel we shall introduce the short-hand notation
$$
z_{\mu}^i:=x^i-v^i_{\mu},\qquad\quad v^i_{\mu}:=
\sum\limits_{\nu=1}^{\mu}y^i_{\nu}, \qquad\qquad \mu=1,2,...,n;
$$
$v^i_{\mu}$ will play the role of coordinates of the center of the
$\mu$-th instanton.
It is easy to check from (\ref{yyrel})
that these new $n$ sets of variables generate as many
copies of the quantum Euclidean space $\b{R}_q^4$, namely
\be
\PH_A{}^{ij}_{hk}z_{\mu}^hz_{\mu}^k=0,  \qquad\quad
\Leftrightarrow \qquad\quad z_{\mu} \bar z_{\mu}=
\bar z_{\mu}z_{\mu}=|z_{\mu}|^2I_2          \label{zzrel}
\ee
and together with $x^i$ make up an alternative
Poincar\'e-Birkhoff-Witt basis of the algebra $\A_n$,
(i.e. ordered monomials in these variables
make up a basis of the vector space underlying $\A_n$).
Moreover, differentiating $z_{\mu}^j$ and commuting it with $\xi^j$ is like
differentiating and commuting $x^j$:
$$
\partial_iz_{\mu}^j=\delta^i_j+q\RH^{jh}_{ik}z_{\mu}^k\partial_h,
\eqno{(\ref{dxrel})}_{\mu}
$$
$$
z_{\mu}^h\xi^i=q\RH^{hi}_{jk}\xi^jz_{\mu}^k. \eqno{(\ref{xxirel})}_{\mu}
$$
Therefore for any $\mu=1,2,...,n$ the replacement $x\to z_\mu$ in any true
relation involving $x,\partial,\xi$ will generate a new true relation,
which we shall label by adding
the subscript $\mu$ to the original one, as we have just done.

The solution $\phi$ searched for (\ref{Dphi}) is of the form
\be
\phi\equiv\phi_n=1+\rho_1^2\frac 1{|x\!-\!y_1|^2}+
\rho_2^2\frac 1{|x\!-\!y_1\!-\!y_2|^2}+...+\rho_n^2\frac
1{|x\!-\!y_1\!-\!...\!-\!y_n|^2}\quad    \label{genphi}
\ee
or, more compactly,
$$
\phi_n=1+\sum\limits_{\mu=1}^n\rho_{\mu}^2\frac 1{|z_{\mu}|^2},
$$
namely a scalar
``function'' of the coordinates $x^i$, of the instanton ``sizes'' $\rho_{\mu}$
and of the ``coordinates of their centers''.
For this to be allowed we have further enlarged $\A_n,\Omega^*(\A_n),{\cal
H}_n,{\cal DC}(\A_n)$ to  extended algebras
$\A^{ext}_n,\Omega^*(\A^{ext}_n){\cal H}^{ext}_n,{\cal DC}(\A^{ext}_n)$   by
adding as generators inverse elements
$1/|z_{\mu}|$, but we also add
the inverses $1/\phi_m$, together with corresponding commutation relations
(see the appendix) consistent with the ones given so far.

By Remark 1 and relation (\ref{utile4}), $\phi$ is harmonic, exactly as in the
classical case.  In the appendix we prove more:

\begin{lemma} Denoting $\phi_q(\{z_i\}):=\phi(\{qz_i\})$,
\bea
&&\Box\phi\sim\bar\partial\partial \phi=\partial\bar\partial \phi=0\qquad
\mbox{(i.e. $\phi$ is harmonic)},               \label{harmonic}\\
&&\phi\xi^i=\xi^i\phi_q,              \label{lulu1}\\
&&[\phi,(\partial_i\phi)]=0=[\phi,(\partial_h\partial_i\phi)],   \label{lulu1'}\\
&&\PH_A{}^{ij}_{hk}(\partial^h\phi)(\partial^k\phi)=0,\label{lulu2}\\
&&(d\phi)(d\phi)=0.               \label{amblimblo1}
\eea
\label{lemma1}
\end{lemma}
We are now ready to prove
\begin{theorem} $\hat A= \big(\hat{\cal D}\phi\big)\phi^{-1}$
with $\phi$ defined in (\ref{genphi}) fulfills the selfduality
equation (\ref{ASDEq})$_1$.
\end{theorem}

\bp~
We denote  $n_q:=1\!+\!q\!+\!...\!+\!q^{n\!-\!1}$.
We find
\bea
&&d\bar\xi=-\bar\xi d
\stackrel{(\ref{xixif})}{=}\frac {-q^2}{1\!+\!q^2}\left[
\epsilon^{-1}\!(\xi\bar\xi\partial)^T\!\epsilon\!+\!
\bar\xi\xi\bar\partial\,\right], \label{gigrobot0}\\
&&\hat{\cal D}\bar\xi\stackrel{(\ref{Doper})}{=} q^3\bar\xi\partial\bar\xi
-\frac {q}{q\!+\!1}d\bar\xi
\stackrel{(\ref{dblu})}{=}-q\bar\xi\xi\bar\partial
-\frac {q^{-1}3_q}{2_q}d\bar\xi, \label{gigrobot1}\\
&&d\bar\xi\partial\phi=-\bar\xi d\partial\phi
\stackrel{(\ref{gigrobot0}),(\ref{harmonic})}{=}\frac {-q^2}{1\!+\!q^2}
\epsilon^{-1}\!(\xi\bar\xi\partial)^T\!\epsilon\partial\phi. \label{gigrobot2}
 \eea
Moreover,
\bea
(\bar\xi\partial\phi)(d\phi)&=&-d\left[(\bar\xi\partial\phi)\phi
\right]+(d\bar\xi\partial\phi)\phi\nn
& \stackrel{(\ref{lulu1}),(\ref{lulu1'})}{=}&
-d\left[\phi_{q^{-1}}(\bar\xi\partial\phi)\right]+(d\bar\xi\partial\phi)\phi\nn
&=&-(d\phi_{q^{-1}}\!)(\bar\xi\partial\phi)\!-\!\phi_{q^{-1}}\!(d\bar\xi\partial\phi)
\!+\!(d\bar\xi\partial\phi)\phi\nn
 & \stackrel{(\ref{lulu1}),(\ref{lulu1'})}{=}&-(d\bar\xi\phi)(\partial\phi)
+(d\bar\xi\partial\phi)(\phi\!-\!\phi_q).\qquad\label{gringo}
\eea
Therefore
\bea
&&\hat F
\stackrel{(\ref{defFgen})}{=} d\left[\big(\hat{\cal D}\phi\big)\phi^{-1}
\right] + \big(\hat{\cal D}\phi\big)\phi^{-1} \big(\hat{\cal
D}\phi\big)\phi^{-1}\nn && \quad = \:\:
\big(d\hat{\cal D}\phi\big) \phi^{-1}+\big(\hat{\cal
D}\phi\big)\phi^{-1}(d\phi)\,\phi^{-1}
+\big(\hat{\cal
D}\phi\big)\phi^{-1}\big(\hat{\cal D}\phi\big)\phi^{-1} \nn && \quad \stackrel{(\ref{lulu1})}{=}
\big(d\hat{\cal D}\phi\big)\, \phi^{-1}+\big(\hat{\cal D}\phi\big)
\left[\big(\hat{\cal D}+d\big)\phi\right]
\phi^{-1} \phi^{-1}_q\nn && \quad \stackrel{(\ref{Doper})}{=}
\big(q^3 d\bar\xi\partial\phi\big) \phi^{-1}+
\big(\hat{\cal D}\phi\big)
\left[\left(q^3\bar\xi\partial+\frac
{1}{q\!+\!1}d\right)\!\phi\right] \phi^{-1} \phi^{-1}_q\nn&& \quad  \stackrel{(\ref{lulu1})}{=}
\big(q^3 d\bar\xi\partial\phi\big) \phi^{-1}+
\left[q^3\big(\hat{\cal D}\bar\xi\phi_q\big)\partial\phi+\frac
{1}{q\!+\!1}\big(\hat{\cal
D}\phi\big)\!d\phi\right] \phi^{-1} \phi^{-1}_q\nn && \quad
\stackrel{(\ref{amblimblo1})(\ref{Doper})}{=}\big(q^3 d\bar\xi\partial\phi\big)
\phi^{-1}+\left\{ q\left(\hat{\cal D}\bar\xi\phi\right)(\partial \phi)
+\frac {q^3}{q\!+\!1}\left(\bar\xi\partial\phi\right)\!(d\phi)\right\}
\phi^{-1} \phi^{-1}_q \nn &&\stackrel{(\ref{gigrobot1}),(\ref{gringo})}{=}
\big(q^3 d\bar\xi\partial\phi\big) \phi^{-1}-\left\{\!
\left[\!\left(\!q^2\bar\xi\xi\bar\partial
\!+\!\frac {3_q}{2_q}d\bar\xi\right)\!\phi\right]\!(\partial \phi)
\right.\nn &&
\left.\qquad\qquad\qquad\qquad\qquad\qquad +\frac {q^3}{2_q}\![(d\bar\xi\phi)(\partial\phi)
\!-\!(d\bar\xi\partial\phi)(\phi\!-\!\phi_q)]\right\}
\phi^{-1} \phi^{-1}_q \nn && =\!q^3
\big(d\bar\xi\partial\phi\big)\! \left[\phi^{-1}\!\!+\!\frac 1{2_q}
(\phi^{-1}_q\!\!-\!\phi^{-1})\!\right]\!-\!\left[q^2\bar\xi
\xi(\bar\partial\phi)(\partial\phi)\!+\!
(q^2\!\!+\!1)(d\bar\xi\phi)(\partial \phi)\right] \!\phi^{-1}_q
\!\phi^{-1} \nn &&\stackrel{(\ref{gigrobot0}),(\ref{gigrobot2})}{=} \frac
{-q^5}{4_q}\left[\epsilon^{-1}\!(\xi\bar\xi\partial)^T\!\epsilon
\partial\phi\right]\left[q\phi^{-1}
\!\!+\!\phi^{-1}_q\right]+q^2\epsilon^{-1}\!(
\xi\bar\xi\partial\phi)^T\!\epsilon(\partial \phi)\phi^{-1} \phi^{-1}_q
;\nonumber\eea
this is a selfdual  matrix, $\hat F\in
M_2\left(\check\Omega^2(\A_n)\right)$,  because $\xi\bar\xi$ is. \ep

Formally, as $x\to \infty$ also $z_{\mu}\to\infty$, $\phi\to 1$,
and a simple inspection shows that
$\hat A\to 0$ as $1/|x|^3$, $\hat F\to 0$ as $1/|x|^4$, exactly
as in the case $q=1$. Therefore $\hat F\hat F$ decreases
fast enough at infinity for integrals like $\int\mbox{tr}(\hat F\hat F)$
to be well defined at infinity.

On the other hand, as $z_{\mu}\to 0$ the function $\phi$
and therefore the gauge potential $\hat A$  are singular,
i.e. formally diverge. We don't know yet whether
the singularity will cause problems also in a proper
functional-analytical treatment (this requires analyzing
representations of the algebra).
If this is the case then, as in the undeformed theory,
the question arises if this
singularity is only due to the choice of a singular gauge and can be removed
by performing a suitable gauge transformation, or it really
affects the field strength. Here we address this issue semi-heuristically. We
shall say that an element of our algebra is: 1. analytic in $z_{\mu}$ if its
power expansion has no poles in $z_{\mu}$, i.e. does not depend on
$1/|z_{\mu}|$; 2. regular in $z_{\mu}$ if it formally keeps finite as
$z_{\mu}\to 0$, i.e. in its power expansion  the
dependence on  $1/|z_{\mu}|$ occurs only through $z_{\mu}/|z_{\mu}|$.
Since such dependences might change upon changing
the order in which the variables $z_1,z_2,...,z_n$, and possible
extra variables $1/|z_1-z_2|,1/|z_1-z_3|,...$ (if necessary),
are displayed, these conditions have to be met for any order.
In appendix \ref{singaugetrans} we show that performing the ``singular
gauge transformation'' $U_2$ defined by
\be
U_2\equiv U_2(z_1,z_2):=\frac{\bar z_1}{|z_1|}\frac{y_2}{|y_2|}
\frac{\bar z_2}{|z_2|}                      \label{defU_2}
\ee
on $\hat A_2$ we obtain a $2$-istanton solution
\be
A_2=U_2^{-1}\left(\hat A U_2+dU_2\right)         \label{A_2}
\ee
analytic in both $z_1,z_2$;  the corresponding  selfdual field
strength will be analytic as well. The form of $U_2$ exactly mimics the
undeformed one of Ref. \cite{GiaRot77,OliSciCre79}. Of course, for
this to make sense, we have to further enlarge the algebras adding
as a generator $1/|y_2|$ with consistent commutation relations; this is done in
the subsection \ref{Addrel}.
By generalization of the undeformed results \cite{GiaRot77,OliSciCre79}, we
are led to the

\medskip
{\bf Conjecture.}
Performing the singular
gauge transformation $U_n$ recursively defined by $U_0=\1_2$ and
\be
U_n\equiv U_n(z_1,...,z_n):=U_{n\!-\!1}(z_1,...,z_{n\!-\!1})
U^{-1}_{n\!-\!1}(y)\frac{\bar z_n}{|z_n|},            \label{defU}
\ee
with $U_{m}(y)$ the function of $y_1,...y_m$ only
defined by $U_m(y):=U_m(z_1\!-\!z_n,...,z_{n\!-\!1}\!-\!z_n)$, we finally obtain a
regular $n$-istanton solution
\be
A\equiv A_n=U_n^{-1}\left(\hat A U_n+dU_n\right)         \label{A_n}
\ee
and a corresponding regular selfdual field strength, for any $n$.

\bigskip
Results for the {\bf $n$-antiinstanton solutions} are obtained
by the already mentioned replacements.
In particular, the singular ones  $\hat A$
are simply obtained replacing $\hat{\cal D}$ with  $\hat{\cal D}'$
in (\ref{Dphi}).

\app{Appendix}
\subsection{Additional relations for the extended algebra}
\label{Addrel}

Let $z:=x\!-\!y$, where $y$ is defined as in section \ref{shift}.
Let $a\cdot b:=a^{\alpha\alpha'}b^{\beta\beta'}
\epsilon_{\alpha\beta}\epsilon_{\alpha'\beta'}$. The following
relations are consequences of the commutation relations for the
generators $x^i,y^i,z^i,\rho_x$ or are (the only) consistent
extensions of these consequences to the square root, inverse, and
inverse square root of $|z|^2,|x|^2, |y|^2$ having the desired, commutative
$q\to 1$ limit.
\bea
&& C\xi^i =q^2 \xi^i C, \qquad \qquad \mbox{for
}\qquad C=|x|^2,|y|^2, y\!\cdot\! x
\,,\,x\!\cdot\! y,|z|^2\label{basic0}\\[8pt]
&& \gamma\xi^i =q^{-1} \xi^i \gamma, \qquad \qquad \mbox{for }\qquad \gamma=
\frac 1{|x|},\frac 1{|y|}, \frac 1{|z|} \label{basic0'}\\[8pt]
&& y^i |x|^2=q^2 |x|^2 y^i, \qquad x^i
|y|^2=q^{-2}|y|^2 x^i, \qquad x\!\cdot\! y=q^2 y\!\cdot\! x,
\label{basic1}\\[8pt] && y^i |x|^{\pm 1}\!=\!q^{\mp 1}|x|^{\pm 1}y^i, \qquad
x^i|y|^{\pm 1} \!=\!q^{\pm 1}|y|^{\pm 1}  x^i, \qquad \frac 1{|y|}\frac 1{|x|}
\!=\!\frac q{|x|}\frac 1{|y|}, \qquad\qquad\label{basic1'}\\[8pt]
&& [y^i, y\!\cdot\! x]= |y|^2 x^i (1-q^{-2}), \qquad
[x^i, y\!\cdot\! x]= -y^i |x|^2 (1-q^{-2}), \label{basic2}\\[8pt]
&& z^i \frac {x^j}{|x|^2}=q^{-1}\RH^{ij}_{hk}\frac
{x^h}{|x|^2}z^k+(1\!-\!q^{-2})
g^{ij},\label{basic1"'}\\[8pt]
&& z \frac {\bar x}{|x|^2}=-q^{-2}\frac{x}{|x|^2}\bar z+
q^{-4}\frac{x}{|x|^2}\cdot z +(1\!-\!q^{-4}),\label{basic1"}\\[8pt]
&& \frac {x^i}{|x|^2}|z|^2\!=\!q^2\!|z|^2\frac {x^i}{|x|^2}\!+\!
(1\!-\!q^2)z^i,\label{conseq1}\\[8pt]
&&\frac 1{|z|^2} \frac{x^i}{|x|^2}\!=\!\frac {x^i}{|x|^2}
\frac {q^2}{|z|^2}\!+\!(1\!-\!q^2)\frac{z^i}{|z|^4},\qquad
\label{conseq2}\\[8pt]
&&\frac {x^i}{|x|^2}|z|=q|z|\frac {x^i}{|x|^2}
\!+\!(1-q)\frac {z^i}{|z|},\label{conseq3}\\[8pt]
&&\frac 1{|z|}
\frac{x^i}{|x|^2}=\frac {x^i}{|x|^2}\frac q{|z|}
\!+\!(1\!-\!q)\frac{z^i}{|z|^3}, \qquad\quad
\frac q{|z|}
\frac{y^i}{|y|^2}=\frac {y^i}{|y|^2}\frac 1{|z|}
\!+\!(1\!-\!q)\frac{z^i}{|z|^3} \label{conseq4}\\[8pt]
&&|z|^2 |x|^2\!=\!|x|^2\!\left[q^4|z|^2\!+\!
(1\!-\!q^2)x\!\cdot\! z\right],\qquad\label{nuova1}\\[8pt]
&&|z|^2 \frac 1{|x|^2}\!=\!\frac {q^{-4}}{|x|^2}|z|^2\!+\!
(1\!-\!q^{-2})\left[q^{-4}\frac
x{|x|^2}\!\cdot\!z\!+\!(1\!-\!q^{-4})\right]\qquad
\label{nuova2}\\[8pt]
&& \frac 1{|z|} \frac 1{|x|^2}\!=\!\frac {q^2}{|x|^2}\frac
1{|z|}\!+\! (q^{-1}\!-\!1)\frac x{|x|^2}\!\cdot\! \frac
z{|z|^3}\qquad \label{nuova5} \\[8pt]
&& \left[\xi^h\frac{x^i}{|x|^2},|z|^2\right]=
(1\!-\!q^2)\xi^h z^i,              \label{furba1}\\[8pt]
&& \left[\xi^h\frac {x^i}{|x|^2},\frac 1{|z|}\right]=
(1\!-\!q^{-1})\xi^h \frac {z^i}{|z|^3} , \qquad         \label{furba2}\\[8pt]
&& \left[Td\overline{T},\frac 1{|z|}\right]\!=\!
(1\!-\!q^{-1})T_zd\overline{T_z}\frac 1{|z|}  .
\label{furba3} \\[8pt]
&& \frac {|x|}{|\rho_x|}\: \mbox{ commutes with }x^i,y^i, z^i, |x|,|y|,|z|
\label{conseq'}
 \eea
For instance, relations (\ref{basic0'})  are postulated by
 consistency with (\ref{basic0}). Relation
(\ref{basic1"}) follows from (\ref{basic1'}), (\ref{xxrel}),
(\ref{Pt}), (\ref{projectorR}), (\ref{yrel})$_1$.
Eq. (\ref{conseq1}), (\ref{conseq2}) follow from the preceding ones.
Relations
(\ref{conseq3}), (\ref{conseq4})  are postulated by
 consistency with (\ref{basic1"}),
(\ref{conseq1}). Eq. (\ref{furba1}) follows from
(\ref{conseq1}),(\ref{conseq2}) and (\ref{basic0}). Relation
(\ref{furba2}) follows from (\ref{conseq3}), (\ref{conseq4}) and
(\ref{basic0'}). Eq. (\ref{furba3}), where we have set $T_z:=z/|z|$,
is a particular consequence of (\ref{furba2}).
 Eq.(\ref{conseq'}) follows from (\ref{new3}).

From (\ref{basic1'}) it also follows
\bea
&&\frac 1{|x|}z=z\frac
q{|x|}+(1\!-\!q)\frac {x}{|x|},\qquad \frac 1{|y|}z=z\frac
{q^{-1}}{|y|}+(q^{-1}\!-\!1)\frac {y}{|y|}\label{lala3} \\&&|x|\frac
1{|y|}x=q^{-1}x|x|\frac 1{|y|},\qquad\qquad\quad
|x|\frac 1{|y|}y=q^{-1}y|x|\frac 1{|y|}, \nn
&&|x|\frac 1{|y|}z\!=\!z|x|\frac {q^{-1}}{|y|}, \qquad|x|\frac
1{|y|}\frac 1{|z|}\!=\!\frac q{|z|}|x|\frac 1{|y|},
\qquad\left[|x|\frac 1{|y|}, \frac {z}{|z|}\right]\!=\!0. \qquad
\label{lala2} \eea

\begin{lemma}
\bea
&& z\frac{\bar y}{|y|}\frac{ x}{|x|}
=\frac{ x}{|x|}\frac{\bar y}{|y|}  z,\qquad\qquad
\bar z\frac{y}{|y|}\frac{\bar  x}{|x|}
=\frac{\bar x}{|x|}\frac{y}{|y|}  \bar z,\label{dindon1}\\[8pt]
&&\bar z\frac{ x}{|x|}\frac{\bar y}{|y|}=
 \frac{\bar y}{|y|}\frac{ x}{|x|}  \bar z,\qquad\qquad
z\frac{\bar x}{|x|}\frac{y}{|y|}=\frac{y}{|y|}\frac{\bar  x}{|x|} z .
\eea
\end{lemma}
\bp~ We use (\ref{lala2})  and (\ref{conseq1})
\be
\bar z\frac{y}{|y|}\frac{\bar x}{|x|} \stackrel{(\ref{basic1'})}{=}
\bar z y\frac{\bar x}{|x|}\frac{1}{|y|}
=\bar z(x\!-\!z)\frac{\bar x}{|x|} \frac 1{|y|}
\stackrel{(\ref{blu})}{=}
\left[\bar z|x|\!-\!|z|^2\frac{\bar x}{|x|}\right]
\frac 1{|y|},    \quad       \label{wow2}
\ee
whereas
\bea
&&\frac{\bar x}{|x|}\frac{y}{|y|} \bar z
=\frac{\bar x}{|x|}\frac 1{|y|}(x\!-\!z)
\bar z\stackrel{(\ref{basic1'})}{=}\frac{\bar
x}{|x|} \left( x\frac {q^{\!-\!1}}{|y|}\!-\!\frac 1{|y|}z\right)\bar z\nn
&& \qquad\stackrel{(\ref{blu}),(\ref{zzrel})}{=}
\left(|x| \frac {q^{\!-\!1}}{|y|}\bar z
\!-\!\frac{\bar x}{|x|}\frac 1{|y|}|z|^2\right)
\stackrel{(\ref{lala2})}{=} q^{\!-\!2}\left(\bar z
\!-\!\frac{\bar x}{|x|^2}|z|^2\right)|x|\frac {1}{|y|}\nn
&&\qquad\stackrel{(\ref{conseq1})}{=}
\left(\bar z|x|
\!-\!|z|^2\frac{\bar x}{|x|}\right)
\frac {1}{|y|}
=rhs(\ref{wow2}).\nonumber
\eea
Completely analogous is the proof of the other relations. \ep

\begin{prop}
\bea
&&\left[|z|^2,\frac{\bar y}{|y|}\frac{ x}{|x|}\right]=
\left[|z|^2,\frac{y}{|y|}\frac{\bar x}{|x|}\right]=
\left[|z|^2,\frac{ x}{|x|}\frac{\bar y}{|y|}\right]=
\left[|z|^2,\frac{\bar x}{|x|}\frac{ y}{|y|}\right]= 0,\label{dundun1}\\[8pt]
&&\left[|z|^{\pm 1}\!,\frac{\bar y}{|y|}\frac{ x}{|x|}\right]\!=\!
\left[|z|^{\pm 1}\!,\frac{y}{|y|}\frac{\bar x}{|x|}\right]\!=\!
\left[|z|^{\pm 1}\!,\frac{ x}{|x|}\frac{\bar y}{|y|}\right]\!=\!
\left[|z|^{\pm 1}\!,\frac{\bar x}{|x|}\frac{y}{|y|}\right]\!=\! 0,
\label{dundun1'}\qquad\quad\\[8pt]
&&U_2(x,z):=\frac{\bar z}{|z|}\frac{y}{|y|}\frac{\bar x}{|x|}=
\frac{\bar x}{|x|}\frac{y}{|y|} \frac{\bar z}{|z|} =\left(\frac{\bar
z}{|z|^2}\!-\!\frac{\bar x}{|x|^2}\right)|x|\frac 1{|y|}|z|,
\label{dundun2}\\[8pt]
&&U^{-1}_2(x,z)=\frac{z}{|z|}\frac{\bar y}{|y|}\frac{ x}{|x|}
=\frac{ x}{|x|}\frac{\bar y}{|y|} \frac{ z}{|z|}=
\left(\frac{ z}{|z|^2}-\frac{ x}{|x|^2}\right)|x|\frac 1{|y|}|z|,\label{dundun3}
\\[8pt]
&&\left[|z|^{\pm 1},U_2(x,z)\right]=0,
\label{dundun6}
\eea
\end{prop}
\bp~ Eq. (\ref{dundun1}) are direct consequences of $|z|^2=\bar
zz=z\bar z$ and of the relations in the lemma. (\ref{dundun1'}) are
derived by consistency with (\ref{dundun1}). The first equality in
(\ref{dundun2}) is a direct consequence of (\ref{dindon1})$_2$ and
of (\ref{dundun1'}); the second equality is a consequence of
(\ref{wow2}), (\ref{conseq3}), (\ref{lala2}).
Eq. (\ref{dundun6}) follows from (\ref{dundun1'}) and
$[z,|z|]=[\bar z,|z|]=0$. \ep

Relations (\ref{dxirel}), (\ref{ymuxirel}),
(\ref{xxirel}), (\ref{yrel})$_2$  respectively imply
\bea
&&\xi\bar\partial+q^2\partial\bar\xi=q^{-2}dI_2,\qquad\qquad\qquad
\bar\xi\partial+q^2\bar\partial\xi=q^{-2}dI_2,  \qquad\quad \label{dblu}\\[8pt]
&&\xi\bar y+y\bar\xi=q^{-2}(\xi\!\cdot\! y)I_2,\qquad\qquad
\bar\xi y+\bar y\xi=q^{-2}\left(\xi\!\cdot\! y\right)I_2,  \qquad\quad  \label{yblu}\\[8pt]
&&\xi\bar z+z\bar\xi=q^{-2}(\xi\!\cdot\! z)I_2,\qquad\qquad
\bar\xi z+\bar z\xi=q^{-2}\left(\xi\!\cdot\! z\right)I_2,  \qquad\quad  \label{zblu}\\[8pt]
&&x\bar y+y\bar x=q^{-2}(x\!\cdot\! y)I_2,\qquad\qquad
\bar x y+\bar y x=q^{-2}\left(x\!\cdot\! y\right)I_2, \qquad \quad  \label{ybluu}
\eea
A quick way to prove these relations is to note that they can be
obtained from (\ref{xblu}) by the following replacements:
$x/|x|^2q^2(1\!-\!q^2)\to\partial$
(see Remark 1),   $x\to y$, $x\to z$, $x\to y$ and $\xi\to x$,
respectively.

\bigskip
{\bf Remark 2.}
By (\ref{lala3}), reordering $1/|x|, 1/|y|$ w.r.t. $x^i,z^i$ does not introduce
additional powers of $1/|x|, 1/|y|,1/|z|$. Consequently, for any $f(x,z)$
analytic w.r.t. $x,z$, $f\,1/|x|=(1/|x|)g$, $f\,1/|y|=(1/|y|)h$, with
$g(x,z)$, $h(x,z)$ analytic  functions w.r.t. $x,z$.
By (\ref{nuova5}), reordering $1/|z|$ w.r.t. $1/|x|^2$ does not introduce
additional powers of $1/|x|$.

\bigskip
{\bf Remark 3.}
Any relation (...) or Remark proved/postulated so far in this appendix is mapped
into a new
true/consistent one, which we shall label as (...)$_{\mu}$ or (...)$_{\mu\nu}$
according to the cases, by the
replacements $x\to z_{\mu}$,  $\rho_x\to \rho_{\mu}$, $y\to
\sum\limits_{\lambda=\mu\!+\!1}^{\nu}y_{\lambda}$, $z\to z_{\nu}$, $\rho_z\to \rho_{\nu}$
with $\nu>\mu$.

\subsection{Proof of Lemma \ref{lemma1}}

Relation (\ref{harmonic}) is a straightforward consequence of
(\ref{new1})$_2$ and of (\ref{utile4})$_{\mu}$, $\mu=1,2,...,n$.
Relation (\ref{lulu1}) is a straightforward consequence of
(\ref{new1})$_1$ and of (\ref{basic0'})$_{\mu}$. To prove
(\ref{lulu1'}), (\ref{lulu2})  we first state the following
relations:
\bea
&&\frac 1{|z_{\nu}|^2}y^i_{\mu}=y^i_{\mu}\frac{q^2}
{|z_{\nu}|^2}, \qquad \frac {\rho_{\nu}^2}{|z_{\nu}|^2}
y^i_{\mu}=y^i_{\mu}\frac {\rho_{\nu}^2}{|z_{\nu}|^2},
\qquad[\frac {\rho_{\nu}^2}{|z_{\nu}|^2},z_{\nu}^i]=0\qquad\nu<
\mu,\label{limbo0} \qquad\\
&&\left[\frac{\rho_{\nu}^2}{|z_{\nu}|^2},z_{\mu}^k\right]\!=\!0,
\qquad\left[\frac{\rho_{\nu}^2}{|z_{\nu}|^2},
\frac{z_{\mu}^k}{|z_{\mu}|}\right]\!=\!0, \qquad\left[\frac
{\rho_{\nu}^2}{|z_{\nu}|^2},\rho_{\mu}^2\right]\!=\!0,
\qquad\nu\le\mu\qquad\quad \label{limbo1}\\
&&\left[\frac {\rho_{\nu}^2}{|z_{\nu}|^2},\frac
{\rho_{\mu}^2}{|z_{\mu}|^2}\right]=0, \label{limbo1'}\\
&&\left[\frac
{\rho_{\nu}^2}{|z_{\nu}|^2},\frac{z_{\mu}^k}{|z_{\mu}|^4}
\rho_{\mu}^2\right]=(1-q^2)
\frac{z_{\nu}^k}{|z_{\nu}|^4}\rho_{\nu}^2 \frac
{\rho_{\mu}^2}{|z_{\mu}|^2}\qquad\qquad \nu>\mu \label{limbo2}\\
&&\PH_A{}^{ij}_{hk}z_{\mu}^hz_{\nu}^k=-\PH_A{}^{ij}_{hk}z_{\nu}^hz_{\mu}^k
\label{limbo3}\\
&&|z_{\nu}|^2\PH_A{}^{ij}_{hk}z_{\mu}^hz_{\nu}^k\frac 1{|z_{\mu}|^2}
=-q^2\PH_A{}^{ij}_{hk}z_{\nu}^hz_{\mu}^k\frac
1{|z_{\mu}|^2}|z_{\nu}|^2 \qquad\qquad \nu>\mu\label{limbo4}
\eea
Relations (\ref{limbo0}) follow from (\ref{basic1'})$_{\mu\nu}$ and
(\ref{new3}). The first two relations (\ref{limbo1}) follow from
(\ref{limbo0}), the third from  (\ref{new3}). (\ref{limbo1'}) is an
immediate consequence of (\ref{limbo1}). Relation (\ref{limbo2}) is
a consequence of (\ref{limbo1}) and (\ref{conseq2})$_{\mu\nu}$,
(\ref{limbo3}) a consequence of (\ref{zzrel}), (\ref{yyrel}) and
(\ref{projectorR}), (\ref{limbo4}) a consequence of (\ref{limbo3}),
(\ref{zzrel})$_{\mu}$ (\ref{conseq1})$_{\mu}$.

To prove (\ref{lulu1'})$_1$ we proceed as follows:
\bea
&&\phi(\partial^k\phi) =-q^{-4} \left[1+\sum\limits_{\mu=0}^n \frac
{\rho_{\mu}^2}{|z_{\mu}|^2}\right] \sum\limits_{\nu=0}^n \frac
{z_{\nu}^k}{|z_{\nu}|^4}\rho_{\nu}^2 \nn
&&=(\partial^k\phi)-q^{-4}\left[\sum\limits_{\mu=0}^n \frac
{\rho_{\mu}^2}{|z_{\mu}|^2}\frac
{z_{\mu}^k}{|z_{\mu}|^4}\rho_{\mu}^2 \!+\!
\sum\limits_{\stackrel{\mu,\nu=0}{\nu>\mu }}^n \frac
{\rho_{\mu}^2}{|z_{\mu}|^2}\frac
{z_{\nu}^k}{|z_{\nu}|^4}\rho_{\nu}^2 \!+\!
\sum\limits_{\stackrel{\mu,\nu=0}{\nu<\mu }}^n \frac
{\rho_{\mu}^2}{|z_{\mu}|^2}\frac
{z_{\nu}^k}{|z_{\nu}|^4}\rho_{\nu}^2\right]\nn
&&\stackrel{(\ref{limbo2})}{=}(\partial^k\phi)-q^{-4}\left[\sum\limits_{\mu=0}^n
\frac {z_{\mu}^k}{|z_{\mu}|^4}\rho_{\mu}^2 \frac
{\rho_{\mu}^2}{|z_{\mu}|^2}+
q^2\sum\limits_{\stackrel{\mu,\nu=0}{\nu>\mu }}^n
\frac{z_{\nu}^k}{|z_{\nu}|^4}\rho_{\nu}^2\frac
{\rho_{\mu}^2}{|z_{\mu}|^2} \right.\nn
&&+\left.\sum\limits_{\stackrel{\mu,\nu=0}{\nu<\mu }}^n
\left(\frac{z_{\nu}^k}{|z_{\nu}|^4}\rho_{\nu}^2\frac 1{|z_{\mu}|^2}
\rho_{\mu}^2-qk\frac{z_{\mu}^k}{|z_{\mu}|^4}\rho_{\mu}^2 \frac
{\rho_{\nu}^2}{|z_{\nu}|^2}\right)\right]\nn &&=(\partial^k\phi)
-q^{-4}\sum\limits_{\nu=0}^n \frac
{z_{\nu}^k}{|z_{\nu}|^4}\rho_{\nu}^2
\left[\sum\limits_{\mu=0}^n\frac {\rho_{\mu}^2}{|z_{\mu}|^2}\right]
=(\partial^k\phi)\,\phi\nonumber
\eea
(in the fourth equality we
have used the fact that the second term in the inner bracket is
proportional to and therefore can be put together with the second in
the square bracket). Similar is the proof of  (\ref{lulu1'})$_2$. As
a consequence we have also $[\phi^{-1},(\partial^k\phi)]=0$ and, by
the replacement $z_{\mu}\to qz_{\mu}$,
$[\phi_q{}^{-1},(\partial^k\phi_q)]=0$. We now prove (\ref{lulu2})
\bea
&&\PH_A{}^{ij}_{hk}(\partial^h\phi)(\partial^k\phi)\nn &&\sim
\PH_A{}^{ij}_{hk} \left[\sum\limits_{\mu=0}^n
\frac{z_{\mu}^h}{|z_{\mu}|^4}\rho_{\mu}^2
\frac{z_{\mu}^k}{|z_{\mu}|^4}\rho_{\mu}^2 +
\sum\limits_{\stackrel{\mu,\nu=0}{\nu>\mu }}^n
\frac{z_{\mu}^h}{|z_{\mu}|^4}\rho_{\mu}^2
\frac{z_{\nu}^k}{|z_{\nu}|^4}\rho_{\nu}^2 +
\sum\limits_{\stackrel{\mu,\nu=0}{\nu<\mu }}^n
\frac{z_{\mu}^h}{|z_{\mu}|^4}\rho_{\mu}^2
\frac{z_{\nu}^k}{|z_{\nu}|^4}\rho_{\nu}^2\right]\nn
&&=\PH_A{}^{ij}_{hk} \left[\sum\limits_{\mu=0}^n
\frac{z_{\mu}^hz_{\mu}^k}{|z_{\mu}|^8}\rho_{\mu}^4 +
\sum\limits_{\stackrel{\mu,\nu=0}{\nu>\mu }}^n
\frac{z_{\mu}^h}{|z_{\mu}|^4}\rho_{\mu}^2
\frac{z_{\nu}^k}{|z_{\nu}|^4}\rho_{\nu}^2 +
\sum\limits_{\stackrel{\mu,\nu=0}{\mu<\nu }}^n
\frac{z_{\nu}^h}{|z_{\nu}|^4}\rho_{\nu}^2
\frac{z_{\mu}^k}{|z_{\mu}|^4}\rho_{\mu}^2 \right]\nonumber
\eea
The first term in the square bracket vanishes because of (\ref{zzrel}),
whereas, because of (\ref{limbo1}-\ref{limbo2}), the other two give,
as claimed,
\bea
&&\PH_A{}^{ij}_{hk}(\partial^h\phi)(\partial^k\phi) \sim
\PH_A{}^{ij}_{hk}\sum\limits_{\stackrel{\mu,\nu=0}{\nu>\mu }}^n
\left[\frac 1{|z_{\mu}|^2}z_{\mu}^h z_{\nu}^k\frac
1{|z_{\nu}|^2}\frac {\rho_{\mu}^2}{|z_{\mu}|^2} \frac
{\rho_{\nu}^2}{|z_{\nu}|^2} +\right.\nn && \left.
\frac{z_{\nu}^h}{|z_{\nu}|^2}\left( \frac{z_{\mu}^k}{|z_{\mu}|^4}
\rho_{\mu}^2\frac {\rho_{\nu}^2}{|z_{\nu}|^2}
+(1-q^2)\frac{z_{\nu}^h}{|z_{\nu}|^4}\rho_{\nu}^2 \frac
{\rho_{\mu}^2}{|z_{\mu}|^2}\right)\right]\nn
&&\stackrel{(\ref{limbo3}),(\ref{zzrel})_{\mu}}{=}
\PH_A{}^{ij}_{hk}\sum\limits_{\stackrel{\mu,\nu=0}{\nu>\mu }}^n
\left[-\frac 1{|z_{\mu}|^2}z_{\nu}^h z_{\mu}^k\frac 1{|z_{\nu}|^2}
\frac {\rho_{\mu}^2}{|z_{\mu}|^2}\frac {\rho_{\nu}^2}{|z_{\nu}|^2} +
\frac{z_{\nu}^h}{|z_{\nu}|^2} \frac{z_{\mu}^k}{|z_{\mu}|^4}
\rho_{\mu}^2\frac {\rho_{\nu}^2}{|z_{\nu}|^2}\right]\nn
&&=
\PH_A{}^{ij}_{hk}\sum\limits_{\stackrel{\mu,\nu=0}{\nu>\mu }}^n
\left[-q^2z_{\nu}^h \frac {z_{\mu}^k}{|z_{\mu}|^2}\frac
1{|z_{\nu}|^2}  \!+\! q^2z_{\nu}^h\frac{z_{\mu}^k}{|z_{\mu}|^2}
\frac 1{|z_{\nu}|^2} \right]\frac {\rho_{\mu}^2}{|z_{\mu}|^2}\frac
{\rho_{\nu}^2}{|z_{\nu}|^2}\!=\!0; \nonumber
\eea
in the last equality we have used (\ref{limbo4}),
(\ref{zzrel})$_{\mu}$,
(\ref{conseq1})$_{\mu}$.
Finally,
\bea
&&(d\phi)(d\phi)=(d\phi)(\xi^i\partial_i\phi)\stackrel{(\ref{lulu1})}{=}
(d\xi^i\phi_q)(\partial_i\phi)=-q^{-2}\xi^i(d\phi)(\partial_i\phi)\nn
&&=-q^{-2}\xi^i\xi^j(\partial_j\phi)(\partial_i\phi)
\stackrel{(\ref{xixirel})}{=}-q^{-2}\xi^h\xi^k\PH_A{}^{ij}_{hk}g_{jl}g_{im}
(\partial^l\phi)(\partial^m\phi)\nn
&&=-q^{-2}\xi^h\xi^kg_{hj}g_{ki}\PH_A{}_{lm}^{ij}
(\partial^l\phi)(\partial^m\phi)\stackrel{(\ref{lulu2})}{=}0\nonumber
\eea
proves (\ref{amblimblo1}). In the last but one equality we have used the
property (see e.g.
\cite{FadResTak89,fiothesis}) $[\PH_A, P(g\!\otimes_{\b{C}}\! g)]=0$,
where $P$ denotes the permutation
matrix.

\subsection{Proof of the analyticity of $A_2$}
\label{singaugetrans}

For any quaternion $w$ let $V(w):=w/|w|$. As a consequence,
$V^{-1}(w):=\bar w/|w|$. So $T=V(x)$. We shall use also the
shorter notation $T_n:=V(z_n)$. Having defined $U_2$ as in (\ref{defU_2}),
we find
\be
U^{-1}_2(dU_2)= T_2V^{-1}(y_2)T_1\big(d\overline{T}_1\big)
V(y_2)\bar T_2 +T_2(d\overline{T}_2). \label{lula2}
\ee

From the definition (\ref{genphi}) it follows, for both $\mu=1,2,$
\be
\phi_2{}^{-1}=\frac{|z_{\mu}|^2}{\rho_{\mu}^2}f_{\mu},
\mbox{where $f_{\mu}$ is analytic in $z_{\mu}$}. \label{lula3}
\ee
Using properties (\ref{limbo1'}), (\ref{limbo1})
we immediately find for  $m=1,2$
\be
[\phi_m,T_2]=0 ,\qquad \qquad    [\phi_1,\phi_2]=0,     \label{birillo}
\ee
whereas we find, as consequences of (\ref{conseq'}), (\ref{dundun1'}),
(\ref{new3})
\be
\left[\phi_m,T_1V^{-1}(y_2)\right]=\left[\phi_m,V(y_2)
\overline{T}_1\right]=\left[\phi_m,U_2\right]=0.      \label{wow4}
\ee
Moreover, by straightforward calculations,
\be
(\hat{\cal D}\phi_2)
=-(d\overline{T}_1)T_1\,\frac{\rho_1^2}{|z_1|^2}
-(d\overline{T}_2)T_2\,\frac{\rho_2^2}{|z_2|^2}.      \label{Dphi_2}
\ee

We first show that $A_2$ is an analytic function of $z_1$:
\bea
&& A_2\stackrel{(\ref{A_2}),(\ref{Dphi})}{=}
U_2^{-1}\left[ \big(\hat{\cal D}\phi_2\big)\phi^{-1}_2U_2+dU_2\right]
\nn &&\stackrel{(\ref{Dphi_2})}{=}U_2^{-1}\left[-\sum\limits_{\mu=1}^2
(d\overline{T}_{\mu})T_{\mu}\,\frac{\rho_{\mu}^2} {|z_{\mu}|^2}
\phi^{-1}_2U_2+dU_2\right] \nn &&\stackrel{(\ref{wow4})}{=}U_2^{-1}\left[
-\sum\limits_{\mu=1}^2 (d\overline{T}_{\mu})T_{\mu}U_2\,\frac{\rho_{\mu}^2}
{|z_{\mu}|^2}  \phi^{-1}_2+dU_2\right]
\nn &&\stackrel{(\ref{lula2})}{=} U_2^{-1}\left[-(d\overline{T}_1)T_1\overline{T}_1
V(y_2)\overline{T}_2\frac{\rho_1^2}{|z_1|^2} \phi^{-1}_2+(d\overline{T}_1)
V(y_2)\overline{T}_2\right.\nn && \left. \qquad-
(d\overline{T}_2)T_2\,U_2
\phi^{-1}_2\frac{\rho_2^2}{|z_2|^2}\right]+T_2(d\overline{T}_2) \nn
\nn &&= U_2^{-1}(d\overline{T}_1)V(y_2)\overline{T}_2
\left(\phi_2-\frac{\rho_1^2}{|z_1|^2}\right) \phi^{-1}_2\nn && \qquad-
U_2^{-1}(d\overline{T}_2)T_2\,U_2
\phi^{-1}_2\frac{\rho_2^2}{|z_2|^2}+T_2(d\overline{T}_2) \nn
&&\stackrel{(\ref{lula3})}{=} T_2V^{-1}(y_2)T_1(d\overline{T}_1)V(y_2)\overline{T}_2
\left(1+\frac{\rho_2^2}{|z_2|^2}\right)\frac{|z_1|^2}{\rho_1^2}f_1
\nn && \qquad -
T_1V^{-1}(y_2)T_2(d\overline{T}_2)\,V(y_2)\overline{T}_1\frac{|z_1|^2}{\rho_1^2}f_1
\frac{\rho_2^2}{|z_2|^2}+T_2(d\overline{T}_2)  \nn
&&\stackrel{(\ref{conseq'})_{12},(\ref{basic0'})_1}{=}
T_2V^{-1}(y_2)T_1(d\overline{T}_1)\frac{|z_1|^2}{\rho_1^2}
V(y_2)\overline{T}_2\left(1+\frac{\rho_2^2}{|z_2|^2}\right)f_1 \nn && \qquad -
q^{-1}\frac{z_1}{\rho_1}V^{-1}(y_2)T_2(d\overline{T}_2)\,V(y_2)\frac{\bar
z_1}{\rho_1}f_1 \frac{\rho_2^2}{|z_2|^2}+T_2(d\overline{T}_2).  \nonumber
\eea
Looking at (\ref{dT}) we see that
$T_1(d\overline{T}_1)|z_1|^2/\rho_1^2$ is analytic in $z_1$; the
factors at its left and right also are. The second term is also a product of
analytic factors in $z_1$. Therefore,  by Remark 2,
the first two terms at the rhs are analytic in $z_1$, however we fix the order
of the variables $z_1,z_2,1/|y_2|$. Finally,
the term $T_2 (d\overline{T}_2)$ is independent of $z_1$.
We conclude that $A_2$ is analytic in $z_1$.

We now show that $A_2$ is analytic in $z_2$. We first prove that
$\hat A_2^{\overline{T}_2}:= T_2\big[\hat A_2\overline{T}_2+(d\overline{T}_2)\big]$
is a regular function of $z_2$, more precisely, even analytic.
\bea
&&\hat A_2^{\overline{T}_2}:= T_2\big[\hat A_2\overline{T}_2
+d\overline{T}_2\big]\stackrel{(\ref{Dphi})}{=}T_2\left[\big(\hat{\cal
D}\phi_2\big)\phi^{-1}_2\overline{T}_2 +\big(d\overline{T}_2\big)\right]\nn
&&\stackrel{(\ref{wow4}),(\ref{Dphi_2})}{=}
T_2\left[-(d\overline{T}_2)T_2\overline{T}_2\phi^{-1}_2\frac
{\rho_2^2}{|z_2|^2}+\big(d\overline{T}_2\big)-
(d\overline{T}_1)T_1\overline{T}_2\phi^{-1}_2
\frac{\rho_1^2}{|z_1|^2}\right]\nn
&&\stackrel{(\ref{Tinverse})}{=}T_2(d\overline{T}_2)\phi^{-1}_2\left(\phi_2-\frac
{\rho_2^2}{|z_2|^2}\right) -
\frac{z_2}{|z_2|}(d\overline{T}_1)T_1
\overline{T}_2\phi^{-1}_2\frac{\rho_1^2}{|z_1|^2}\nn
&&\stackrel{(\ref{furba3})_{12}}{=}T_2(d\overline{T}_2)\phi^{-1}_2
\phi_1 -
z_2\left[(d\overline{T}_1)T_1
-(1\!-\!q^{-1})(d\overline{T}_2)T_2 \right]\frac{\overline{T}_2
}{|z_2|}\phi^{-1}_2 \frac{\rho_1^2}{|z_1|^2}\nn
&&\stackrel{(\ref{Tinverse}),(\ref{basic0'})_2}{=}
T_2(d\overline{T}_2)\phi^{-1}_2 \left[\phi_1
+(q\!-\!1)(\phi_1\!-\!1)\right]- z_2(d\overline{T}_1)T_1
\frac{\rho_1^2}{|z_1|^2}\frac{\bar z_2}{|z_2|^2}\phi^{-1}_2 \nn
&&\stackrel{(\ref{lula3})}{=}T_2(d\overline{T}_2)
\frac{|z_2|^2}{\rho_2^2}f_2 \left[q\phi_1
+(1\!-\!q)\right]-z_2(d\overline{T}_1)T_1
\frac{\rho_1^2}{|z_1|^2}\frac{\bar z_2}{\rho_2^2} f_2\nonumber
\eea
As $\phi_1$ does not depend on $z_2$ and
$T_2(d\overline{T}_2)|z_2|^2, f_2$  are analytic in $z_2$, the first term is.
On the other hand,  the
second term is manifestly analytic in $z_2$.
Now, by a further gauge transformation $\tilde U:=V(y_2)\overline{T}_1$,
$$
A_2=\tilde U^{-1}\hat A^{\overline{T}_2}\tilde U+
T_1(d\overline{T}_1).
$$
$\tilde U$ is an analytic function of $z_2$, therefore by Remark 2
the first term remains analytic in $z_2$ (however we fix to
order the variables $z_1,z_2,1/|y_2|$); the second term
is even independent of $z_2$, so $A_2$ is analytic  in $z_2$.


\end{document}